\documentclass[onecolumn,showpacs,amsmath,amssymb]{revtex4}
\usepackage{graphicx}
\usepackage{bm}
\begin{document}
\title{Topological Berry phase of a pumped three-level atom in the presence of dispersing and absorbing  dielectric  bodies: application to a half-space dielectric surface}
%
%
\author{M. S. Ateto\footnote{E-mail: omersog@yahoo.com; mohamed.ali11@sci.svu.edu.eg; msoria@jazan.edu.sa}}
\affiliation{Mathematics Department, Faculty of Science at Qena, South Valley
University, 83523 Qena, Egypt\\
Mathematics Department, Faculty of Science, Gazan University, Gazan, Kingdom of Saudi Arabia}
\begin{abstract}
Within the framework of exact quantum electrodynamics in dielectric, we study the topological Berry phase of a classically pumped $\Lambda$-type three-level atom, prepared initially in a superposition of its two pumped levels and located near a planar dielectric half-space with model permittivity of Lorentz type, interacs with a vacuum field. The effects of material losses, expressed in terms of the Green tensor of the dielectric-matter formation including dispersion and absorption, on the topological Berry phase of the photon transition between the pumped levels have been studied. The outcomes are compared with that of an atom in free space.  We expect, the additional noise due to presence of the dielectric will modify, in a unified way, the behavior and values measured  of the phase shift between the wavefunction evolution and its original state. It is shown that for small separation distance between atom and the dielectric surface, the angle separation between the wavefunction evolution and the original state reduced noticeably that is of strong applications in construction of the universal quantum logic gates. Further, the study opens routes for new challenging applications of atomically systems as various sources of coherent light emitted by pumped atoms in dielectric surroundings.
\end{abstract}
\maketitle
\section{Introduction}
In experiments, the use of optical instruments, which are macroscopic material bodies, such as beam splitters or cavities, requires careful examination with regard to their action on the light under study. In principle, such bodies could be included as a part of the matter to which the radiation field is coupled and treated microscopically \cite{VogelWelsch06}. Dielectric matter plays an important role in optics, because (passive) optical instruments are typically composed of dielectrics. In quantum optics an important consideration is the influence of the presence of material bodies on the quantum statistics of the light. Commonly, dielectric matter is characterized by the permittivity; a complex function of frequency with real part (responsible for dispersion) and imaginary part (responsible for absorption), which describes the response of the matter to the electric field. From statistical mechanics
it is clear that dissipation is unavoidably connected with the appearance of a random force which gives rise to an additional noise source of the electromagnetic field. Hence, any quantum theory that is based on the assumption of a real permittivity can only be valid for narrow-bandwidth fields far from medium resonances where absorption can safely be disregarded. In 1946, Purcell \cite{Purcell46}, through Purcell effect, has stimulated series of theoretical and experimental works . The effect lies in the fact that the spontaneous decay rate of an excited atom essentially depends on whether or not the atom is located near optical inhomogeneities and interfaces of media with differing optical properties. In fact, spontaneous decay, which is a property of spontaneous emission of an excited atom,  depends sensitively on the photonic spectral density of states that are involved in the atomic transition at a chosen location of the atom. However, depending on the specific configuration of inhomogeneities (interfaces); expressed in terms of the Green tensor of the dielectric-matter formation including absorption and dispersion, the atomic spontaneous decay rate may both increase and decrease compared with that of the same atom in free space according to  the modifications in the photonic density of states due to the presence of macroscopic bodies. It is well known that the spontaneous decay of an excited atom can be strongly modified when it is placed inside a microcavity \cite{Berman94, Hinds91}. Purcell effect took on special significance in view of rapid progress in physics of low-dimensional nanostructures. It was shown to be of great importance for microcavities \cite{DKW01}, optical fibers \cite{SOTR01}, photonic crystals \cite{YAZH00}, semiconductor quantum dots \cite{SUGA95}. At this point, a very interesting question is that of the topological phase (i.e., the adiabatic geometric phase) of the wavefunction due to the presence of the matter. Obviousely, the quantum vacuum effects, such as, the Casimir effect \cite{Casimir48, VogelWelsch06}, anomalous magnetic moment of electron,  vacuum polarization leading to Lamb’s shift \cite{BetheBrown47} as well as Casimir-Polder potentials \cite{VogelWelsch06, SpruchKesley78}, atomic spontaneous radiation and its modification in photonic crystals \cite{Yablonovitch87}, magnetoelectric birefringences of quantum vacuum  \cite{RikkenRizzo00} and vacuum contribution to the momentum of anisotropic media (e.g., magnetoelectric materials) \cite{Feigel04}, will strongly influenced due to the presence of these dielectric bodies and as a consequence the change in phase shift between the evolution of the  wavefunction and its original state. 
\par Geometric phases have received special attention of many researchers since Berry showed that there exists a topological phase in quantum-mechanical wavefunction in adiabatic quantum process where the cyclic evolution of wavefunction yields the original state plus a phase shift, which is a sum of a dynamical phase and a geometric phase shift \cite{Berry84}. Berry phase \cite{Berry84} or geometric phase, which does not have classical correspondence, became a focus point in modern physics. It describes a phase factor gained by the wavefunction after the system undergoes an adiabatic and cyclic evolution, which reflects the topological properties \cite{Simon83, SamuelBhardari88} of the state space of the system and has untrivial connections with the character of the system \cite{WangFeng08}, especially with the entanglement \cite{Basu06, CuiWangYi07}. The associated topological Berry’s phase is known as the “Pancharatnam phase” \cite{Panch56}, who introduced the concept of a geometric phase during  his studies of interference effects of polarized light waves. Recently, the Berry phase was introduced into quantum computation to construct a universal quantum logic gates that may be robust to certain kinds of errors \cite{EkertInamori00, WangKeiji01, ZhuWang03, ZhangZhu05, WangWuKwek07}. It was shown \cite{EkErHaInVe00, ZanRas99, PaZan02, Falci00, DuCiZo01, JoVedEk99, GaCi03, WangKeiji01, Solinas03, Solina03, ZhuWan03} that the geometric phase shift can be used for generation of built-in fault-tolerance phase shift gates in quantum computation.  Chen et al., \cite{ChenKanggFeng09} investigated the geometric phase gate in a system with many identical three-level atoms confined in a cavity under decay, driven by a classical field. They \cite{ChenKanggFeng09} proposed an efficient scheme for implementation of two-qubit nonconventional  geometric quantum gates, based on a dissipative large-detuning interaction of two-three-level atoms with a cavity mode is initially in the vacuum state . Many generalizations have been proposed to the original definition \cite{ShWi89,SjPaEkAnEr00, ToSjFiKwOh05, Ben04,LaLaJo99}. The basic theory behind geometric phases and their importance in quantum theory have been reported in \cite{Vedral03, Pati03, ErSjBrOi03, CarolloGuridi04}. In \cite{LiGoCh99} experiments are proposed for the observation of the nonlinearity of the Pancharatnam phase with a Michelson interferometer. Jordan \cite{Jordan88} has used the ideas of Pancharatnam to obtain a definition of phase change for partial cycles. The ideas of Pancharatnam were also used \cite{SamBh88, Berry87} to show that for the appearance of Berry’s phase the system needs to be neither unitary nor cyclic \cite{WeiBan90, WuLi88}, and may be interrupted by quantum measurements. However, this is not every thing, the topics on geometric phases and time-dependent quantum systems \cite{Berry84} still attract  extensive attention of a large number of investigators in various fields, including quantum optics \cite{GongChen99,  YugJiao10}, condensed matter physics \cite{Taguchi01, Falci00}, nuclear physics \cite{Wagh00}, gravity theory \cite{FurtadoBezerra00} as well as molecular physics (molecular chemical reaction) \cite{ShenHe03}.  To the best of our knowledge, the study of the topological Berry phase \cite{Panch56} of an atom-field interaction, within the framework of exact quantum electrodynamics in dispersing and absorbing media, has so far never been considered in references. A much richer range of phenomena is to be expected when allowing for the presence of dispersing and absorbing media, where a complex interplay of the electric properties of the atom and the bodies influences the functional dependence of the atom-field interaction. This problem is addressed in the current work, where we shed light on the topological Berry phase within the framework of the cavity QED, which is an important solid-state system for implementing quantum computation. We expect that, the additional noise due to the presence of dielectric media will, considerably, affect the measured values and behavior of the argument overlap between the wavefunction evolution and its original state. This approach, which has the advantage of being simple and applicable to different configurations of three-level systems, renders general expressions for the three-atom wave vector, as shown in Secs. II and III. In Sec. IV, application to the general results are examined for an atom placed in free space as well as in front of a dielectric half-space. A definition to the topological Berry phase and the form we are going to use including our numerical calculation and discussion are given in Sec. V. A summary is given in Sec. VI.
\section{GENERAL FORMALISM}
Consider a neutral atomic system with its center-of-mass positioned at the point ${\bf r}_A$  near an infinitely long single-wall of dielectric surface. The total nonrelativistic minimal-coupling Hamiltonian of an atomic system interacting with a medium-assisted vacuum electromagnetic field in the form \cite{KSW01}
\begin{equation}
 \label{E1}
\hat{H}=\int d^{3}{\bf r}\int_{0}^{\infty} d\omega~\hbar\omega~ \hat{\bf f}^{\prime\dag}({\bf r},\omega)~\hat{\bf f}^{\prime}({\bf r},\omega) 
+\sum_{j}\frac{1}{2m_{j}}\biggl[\hat{{\bf p}}_{j}- Q_{j}\hat{{\bf A}}(\hat{{\bf r}}_{j})\biggr]^{2}
+\frac{1}{2}\int d^{3} {\bf r}\hat{\rho}_A ({\bf r})\hat{\varphi}_{A}({\bf r})+\int d^{3} {\bf r} \hat{\rho}_A ({\bf r}){ \hat{\varphi}}({\bf r}),
\end{equation}
where $m_j$, $Q_{j}$, ${\bf r}_{j}$ and ${\bf p}_{j}$ are, respectively, the mass, charge, position operator (relative to ${\bf r}_A$)  and  momentum of the $j$th particle constituting the atomic subsystem.
 In the Hamiltonian (\ref{E1}),  the electric field strength is expressed in terms of a continuum set of bosonic fields $\hat{\bf f}({\bf r},\omega)$ and $\hat{\bf f}^{\dag}({\bf r},\omega)$  which play the role of the fundamental (dynamical) variables of the composed system (electromagnetic field and the medium including the dissipative system) and satisfy the well-known commutation relations
\begin{equation}
 \label{E2}
 \big[\hat{\bf f}_{m}({\bf r},\omega),\hat{\bf f}_{n}^{\dag}({\bf r}^{\prime},\omega^{\prime})\big]=\delta_{mn}\delta(\omega-\omega^{\prime})\delta({\bf r}-{\bf r}{}'),
\end{equation}
\begin{equation}
\label{E3}
\big[\hat{\bf f}_{m}({\bf r},\omega),\hat{\bf f}_{n}({\bf r}^{\prime},\omega^{\prime})\big]=\big[\hat{\bf f}^{\dag}_{m}({\bf r},\omega),\hat{\bf f}^{\dag}_{n}({\bf r}^{\prime},\omega^{\prime})\big]=0.
\end{equation}
In terms of the variables $\hat{\bf f}({\bf r},\omega)$ and $\hat{\bf f}^{\dag}({\bf r},\omega)$ the medium-assisted electric field  operator $\hat{{\bf E}}({\bf r})$ can be expressed as
\begin{equation}
\label{E4}
\hat{\bf E}({\bf r})=\hat{\bf E}^{(+)}({\bf r})+\hat{\bf E}^{(-)}({\bf r}),
\end{equation}
with 
\begin{equation}
\label{E5}
\hat{\bf E}^{(+)}({\bf r})=\int_{0}^{\infty}d\omega~\hat{\bf E}({\bf r},\omega),~~~~~~~~~~~ \hat{\bf E}^{(-)}({\bf r})=[\hat{\bf E}^{(+)}({\bf r})]^{\dag},
\end{equation}
\begin{equation}
\label{E6}
\hat{\bf E}({\bf r},\omega)=i\sqrt{\frac{\hbar}{\varepsilon_{0}\pi}}\frac{\omega^{2}}{c^{2}}\int d^{3} {\bf r}{}'\sqrt{\varepsilon_{I}({\bf r}^{\prime},\omega)}~\bm{G}({\bf r},{\bf r}{}',\omega)~.~\hat{\bf f}({\bf r}^{\prime},\omega),
\end{equation}
where $\bm{G}({\bf r},{\bf r}{}',\omega)$ is the classical Green tensor satisfying the equation
\begin{equation}
\label{E7}
 \Big[\frac{\omega^2}{c^2}\varepsilon({\bf r},\omega)-\nabla\times\nabla\times\Big]\bm{G}({\bf r},{\bf r}{}',\omega)=-\delta({\bf r}-{\bf r}{}'),
\end{equation}
and satisfies the boundary condition at infinity, i. e.,
\begin{equation}
\label{E8}
\bm{G}({\bf r},{\bf r}{}',\omega)\rightarrow 0~~\text{if}~~\mid {\bf r}-{\bf r}{}'\mid\rightarrow\infty,
\end{equation}
with complex permittivity $\varepsilon({\bf r},\omega)$ is a function of frequency and space with the real part $\varepsilon_{R}({\bf r},\omega)$ and the imaginary part $\varepsilon_{I}({\bf r},\omega)$ satisfy the Kramers-Kronig relations (for any ${\bf r}$). The second and third terms of the Hamiltonian (\ref{E1}), represent the kinetic energy of the charged particles and their mutual Coulomb interaction, respectively,
\begin{equation}
\label{E9}
 \hat{\varphi}_{A}({\bf r})=\int d{\bf r}^{\prime}\frac{\hat{\rho}_A ({\bf r}^{\prime})}{4\pi\varepsilon_{0}\mid {\bf r}-{\bf r}^{\prime}\mid},
\end{equation}
is the scalar potential of the charged particles distributed with the density,
\begin{equation}
\label{E10}
 \hat{\rho}_A ({\bf r})=\sum_{j} Q_{j}\delta({\bf r}-{\bf r}_{j}),
\end{equation}
in the atomic subsystem. The last term of the Hamiltonian accounts for the Coulomb interaction of the particles with the the medium.
\subsection{Hamiltonian in the electric dipole approximations}
Applying the electric-dipole approximation, we may write the Hamiltonian (\ref{E1}) in the form (see Appendix A) 
\begin{equation}
\label{E11}
\hat H=\int d^3 {\bf r}\int_{0}^{\infty} d\omega~\hbar\omega~ \hat{\bf f}^{\dag}({\bf r},\omega)\cdot\hat{\bf f}({\bf r},\omega)+\sum_{k}\hbar\omega_{k}\hat{S}_{Akk}-\hat{{\bf d}}_{A}\cdot \hat{{\bf E}}({\bf r}_A).
\end{equation}
In this equation, the second term is the Hamiltonian of $k$-level atom, where the $\hat{S}_{Akk}$ are the atomic flip operators, while  the last term is the atom-field coupling energy, where $\hat{{\bf d}}_{A}$ is the electric dipole atomic operator, [see Appendix A, Eqs. (\ref{A9}, \ref{A11})], where the medium-assisted electric field $\hat{\bf E}({\bf r})$ is expressed in terms of the variables $\hat{\bf f}^{\dag}({\bf r},\omega)$ and $\hat{\bf f}({\bf r},\omega)$ according to Eqs.(\ref{E4}-\ref{E8}).
\subsection{ Dynamics of a pumped three-level $\Lambda$-type atom}
Let us now focus on a three-level $\Lambda$-type atom, [Fig. 1(a)], near a one-dimensional infinitely long single-wall dielectric layer - the atom is on the right of the plate, with $z_A$ being the (positive) distance between the surface of the plate and the atom [see Fig. 1(b)]. In particular, we assume that the atomic transition $\mid 2\rangle\rightarrow\mid 3\rangle$ is strongly coupled to the field modes via the dipole  $\hat{\bf d}_{23}$. Moreover, we assume that an external (classical) pump field, with frequency $\omega_{pmp}$ and intensity described by the Rabi frequency $\Omega_{pmp}$, is applied to the atomic transition $\mid 1\rangle\rightarrow\mid 2\rangle$. We may start from the one-dimensional version of the Hamiltonian (\ref{E11}) and apply the rotating-wave approximation [see Appendix A], the Hamiltonian (\ref{E11}) reads
\begin{figure}[tpbh]
\noindent
\begin{center}
\includegraphics[width=0.35\linewidth]
{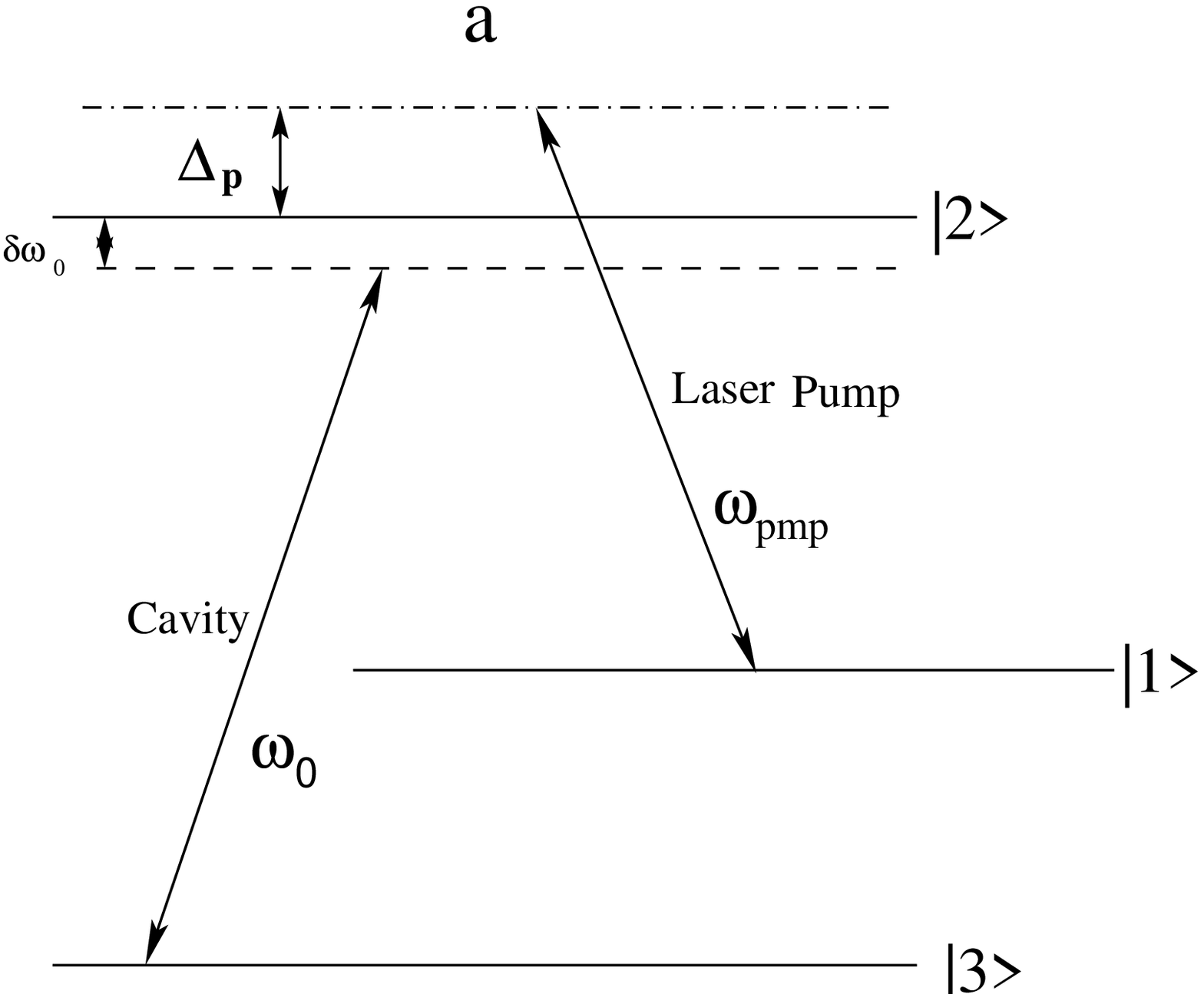}
\hspace{1.5cm}
\includegraphics[width=0.45\linewidth]
{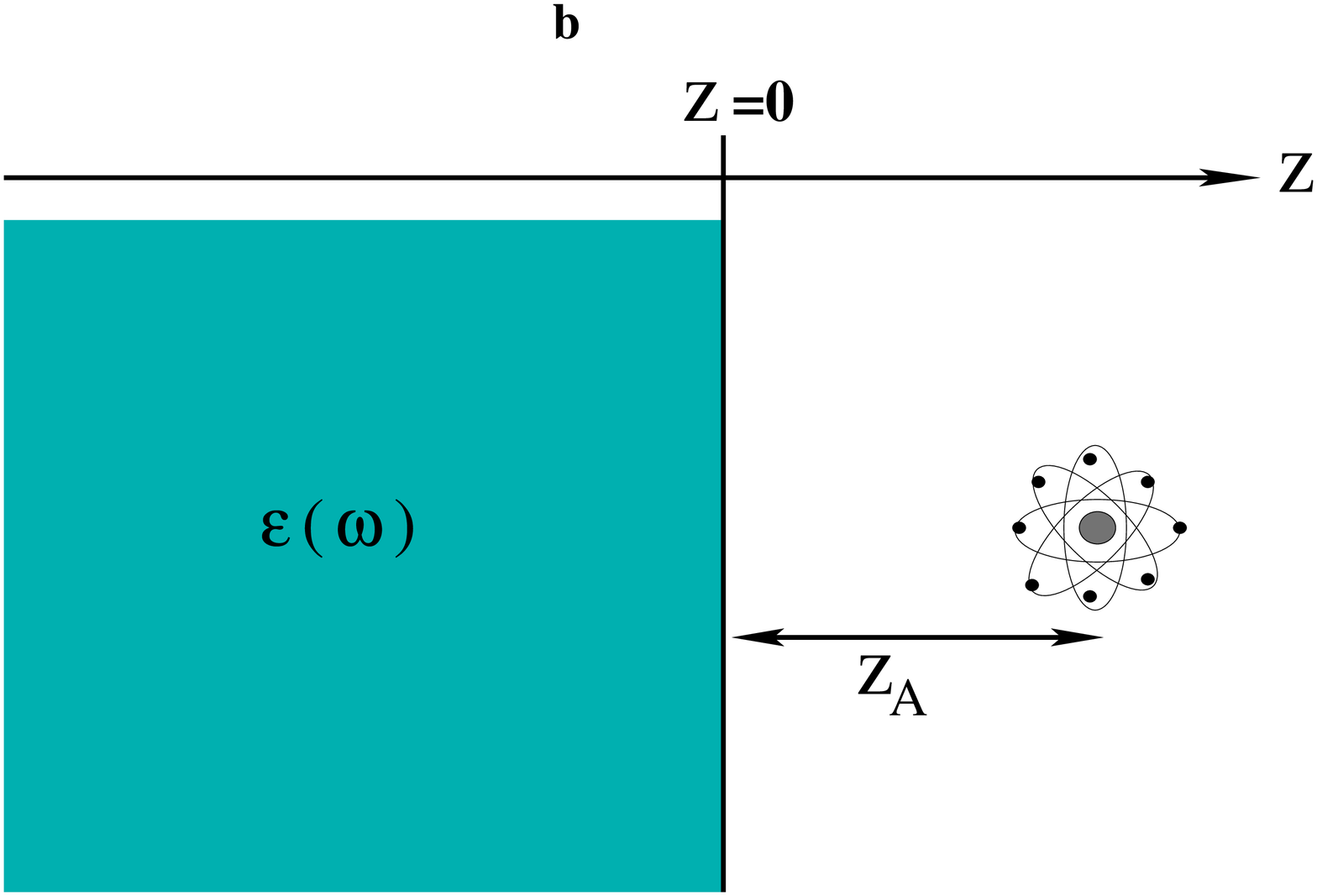}
\end{center}
\caption{(a) Scheme of relevant energy levels and transitions for
the three-level $\Lambda$-type atom, (b) An atom interacting with a half-space dielectric plate with a chosen atomic position $z_A$.}
\label{Fig.1}
\end{figure}
\begin{equation}
\label{E12}
\hat H=\int dz\int_{0}^{\infty} d\omega~\hbar\omega~ \hat{\bf f}^{\dag}({\bf r},\omega)~\hat{\bf f}({\bf r},\omega)
+\hbar\omega_{0}\hat{S}_{22}
-\Big(d_{23}\hat{S}_{23}\hat{E}^{(+)}(z_{A})+\mathrm{H.c.}\Big)
-\frac{\hbar\Omega_{pmp}}{2}\Big(\hat{S}_{21}~e^{-i(\vartheta_{pmp}+\omega_{pmp}t)}+\mathrm{H.c.}\Big).
\end{equation}
\section{Equation of motion}
In what follows we assume that the atom is  initially (at time $t=0$)  prepared in a superposition of its two classically pumped levels $\mid 1\rangle$ and $\mid 2\rangle$, and the rest of the system, that consists of the electromagnetic field and the cavity, is in ground (vacuum) state $\mid\{0\}\rangle$. For simplicity, and without loss of generality, we assume that the frequencies of the atomic transitions $\mid 2\rangle\leftrightarrow\mid 1\rangle$ and $\mid 2\rangle\leftrightarrow\mid 3\rangle$ are the same; $\omega_{21}=\omega_{23}=\omega_{0}$. We may therfore expand the state vector of the overall system at later time $t\ge 0$ as
\begin{equation}
\label{E13}
\mid\psi(t)\rangle=C_{1}(t)\mid 1\rangle\mid\{0\}\rangle+C_{2}(t)~e^{-i\omega_{0}t}\mid 2\rangle\mid\{0\}\rangle
+\int dz\int_{0}^{\infty}d\omega~C_{3}(z,\omega,t)~e^{-i\omega t}\mid 3\rangle~\hat{\bf f}^{\dag}(z,\omega)\mid\{0\}\rangle,
\end{equation}
where  $\hat{\bf f}^{\dag}(z,\omega)\mid\{0\}\rangle$ is a single-quantum excited state of the combined field-cavity system. It is easy to prove that the Schr\"{o}dinger equation for $\mid\psi(t)\rangle$ leads to the following system of (integro-)differential equations for the probability amplitudes $C_{1}(t)$, $C_{2}(t)$ and $C_{3}(z,\omega,t)$:
\begin{equation}
\label{E14}
\dot{C}_{1}(t)=i\frac{\Omega_{pmp}}{2} e^{i\vartheta_{pmp}}~e^{i\Delta_{p} t}C_{2}(t),
\end{equation}
with 
\begin{equation}
\label{E15}
\Delta_{p}=\omega_{pmp}-\omega_{0},
\end{equation}
\begin{equation}
\label{E16}
\dot{C}_{2}(t)=i\frac{\Omega_{pmp}}{2} e^{-i\vartheta_{pmp}}~e^{-i\Delta_{p}t}C_{1}(t)
- \frac{d_{23}}{\sqrt{\hbar\varepsilon_{0}\pi}}\int dz\int_{0}^{\infty} d\omega\frac{\omega^{2}}{c^{2}} \sqrt{\varepsilon^{\prime\prime}(z,\omega)}~\bm{G}(z_{A},z,\omega)~C_{3}(z,\omega,t)~e^{-i(\omega-\omega_{0}) t},
\end{equation}
\begin{equation}
\label{E17}
 \dot{C}_{3}(z,\omega,t)=\frac{d_{32}}{\sqrt{\hbar\varepsilon_{0}\pi}}\frac{\omega^{ 2}}{c^{2}} \sqrt{\varepsilon^{\prime\prime}(z,\omega)}~\bm{G}^{\ast}(z_{A},z,\omega)~C_{2}(t)~e^{i(\omega-\omega_{0}) t}.
\end{equation}
Substituting the formal solution of Eq. (\ref{E17}) [with the initial condition $C_{3} (z,\omega, 0) = 0$]
\begin{equation}
\label{E18}
C_{3}(z,\omega,t)=\frac{d_{32}}{\sqrt{\hbar\varepsilon_{0}\pi}}\frac{\omega^{ 2}}{c^{2}} \sqrt{\varepsilon^{\prime\prime}(z,\omega)}~\bm{G}^{\ast}(z_{A},z,\omega)\int_{0}^{t}~dt^{\prime}~C_{2}(t^{\prime})~e^{i(\omega-\omega_{0}) t^{\prime}},
\end{equation}
and of Eq. (\ref{E14}) [with the initial condition $C_{1} (t=0) =C_{1} (0)$] into Eq.(\ref{E16}), and employing the integral relation
\begin{equation}
\label{E19}
\frac{\omega^{2}}{c^{2}}\int dz \varepsilon^{\prime\prime}(z,\omega)~G(z_{A},z,\omega)G^{\ast}(z_{A},z,\omega)=\mathrm{Im}~ G(z_{A},z_{A},\omega),
\end{equation}
we arrive at
\begin{equation}
\label{E20}
\dot{C}_{2}(t)=i\frac{\Omega_{pmp}(t)}{2} e^{-i\vartheta_{pmp}}~e^{-i\Delta_{p}t}C_{1}(t)
+\int_{0}^{t}~dt^{\prime}~K(t,t^{\prime})~C_{2}(t^{\prime}),
\end{equation}
with the kernel function at the position $z_A$
\begin{equation}
\label{E21}
 K(t,t^{\prime})=-\frac{|d_{23}|^{2}}{\hbar\varepsilon_{0}\pi c^{2}}\int_{0}^{\infty} d\omega~ \omega^{2} \mathrm{Im}~ \bm{G}(z_{A},z_{A},\omega)~e^{-i(\omega-\omega_{0}) (t-t^{\prime})}.
\end{equation}
It is worth noting that  all the matter parameters that are relevant for the atomic evolution are contained, via the Green
tensor, in the kernel function (\ref{E21}).
\section{APPLICATIONS}
\subsection{Three-level atom in free space near a perfectly reflecting mirror}
Besides being an interesting case in its own right, the free-space case will be useful in the comparison and interpretation of the outcomes. Since the electromagnetic vacuum cannot be switched off, its interaction with atomic systems cannot be switched off either, thereby giving rise to a number of observable effects such as spontaneous emission, the Lamb shift, intermolecular energy transfer, and so forth. In free space, for a sufficiently high-Q vacuum cavity contains nondispersing and nonabsorbing media, the excitation spectrum of the electromagnetic field effectively  extends over a frequency interval, so that the $\omega$ integrals [see, e.g., (\ref{E21})] effectively run only over this interval. If we assume that the frequency interval is sufficiently small and sufficiently far from medium resonances, so that both dispersion and absorption may be disregarded. In this case, the permittivity becomes approximately real and independent of frequency. Note that a constant and real permittivity in the whole frequency domain can be realized only for unity permittivity, i.e, the vacuum. This  can be used to express the vector potential and the canonical momentum field in terms of the monochromatic modes and the associated photon annihilation and creation operators.  In short, the excitation spectrum turns into a quasi-discrete set of lines of mid-frequencies $\omega_k$ and widths $\Gamma_k$. Furthermore, in the emission spectrum, the intensity of light associated with the emitted radiation is going to be centered about the atomic transition frequency $\omega_{0}$, with widths $\Gamma_k\approx 0$. 
Bearing in mind these physical interpretations, with the Green tensor $G(z_{A},z_{A},\omega)$ is replaced by the vacuum Green tensor; $G^{V}(z_{A},z_{A},\omega_{0})$ \cite{KSW01}, where 
\begin{equation}
\label{E22}
\mathrm{Im} G^{V}(z_{A},z_{A},\omega_{0})=\frac{\omega_{0}}{6\pi c}\mathbf{I},
\end{equation}
equation (\ref{E21}) becomes
\begin{equation}
\label{E23}
 K(t,t^{\prime})=-\frac{|d_{23}|^{2}}{\hbar\varepsilon_{0}\pi c^{2}}\frac{\omega_{0}}{6\pi c}\int_{0}^{\infty} d\omega_k~ \omega^{2}_k ~e^{-i(\omega_k-\omega_{0}) (t-t^{\prime})}.
\end{equation}
The quantity $\omega^{2}_k$ varies little around $\omega_k\approx\omega_{0}$, we can therefore replace  $\omega^{2}_k$ by  $\omega^{2}_0$. The integral 
\begin{equation}
\label{E24}
\int_{-\infty}^{\infty} d\omega_k~ e^{-i(\omega_k-\omega_{0}) (t-t^{\prime})}=2\pi\delta(t-t^{\prime}).
\end{equation}
yields the following equation for $C_2 (t)$, in the Weisskopf-Wigner approximation
\begin{equation}
\label{E25}
\dot{C}_{2}(t)=i\frac{\Omega_{pmp}}{2} e^{-i\vartheta_{pmp}}~e^{-i\Delta_{p}t}C_{1}(t).
-\Gamma^{32}_{0}C_2 (t)
\end{equation}
with
\begin{equation}
\label{E26}
 \Gamma^{32}_{0}=\frac{|d_{23}|^{2}\omega_{0}^3}{3\hbar\varepsilon_{0}\pi c^{3}}
\end{equation}
 is the free-space spontaneous emission rate for the transition $\mid 2\rangle\rightarrow\mid 3\rangle$.
\par Upon taking the Laplace transforms of equations (\ref{E14}) and (\ref{E25}) with the  use of the initial conditions [$C_{1} (t=0) =C_{1} (0)$ and $C_{2} (t=0) =C_{2} (0)$], we find that
\begin{equation}
\label{E27}
\widetilde{C}_{2}(s)=\frac{(s+i\Delta_{p})C_{2}(0)+i\frac{\Omega_{pmp}}{2}e^{-i\vartheta_{pmp}}C_{1}(0)}{\wp(s)},
\end{equation}
and
\begin{equation}
\label{E28}
\widetilde{C}_{1}(s+i\Delta_{p})=\frac{[s+\Gamma^{32}_{0}]C_{1}(0)+i\frac{\Omega_{pmp}}{2}e^{i\vartheta_{pmp}}C_{2}(0)}{\wp(s)},
\end{equation}
where $\tilde{C}_{j}$ are the Laplace transforms of $C_j$, and
\begin{equation}
 \label{E29}
\wp(s)=s^{2}+\Big[\Gamma^{32}_{0}+i\Delta_{p}\Big]s+\Big[\frac{\Omega_{pmp}^{2}}{4}+i\Gamma^{32}_{0}\Delta_{p}\Big]=\prod_{k=1}^{2}(s-x_{k})
\end{equation}
and $x_{k},(k=1,2)$ are the roots of the quadratic equation
\begin{equation}
 \label{E30}
s^{2}+\Big[\Gamma^{32}_{0}+i\Delta_{p}\Big]s+\Big[\frac{\Omega_{pmp}^{2}}{4}+i\Gamma^{32}_{0}\Delta_{p}\Big]=0,
\end{equation}
found by substituting $x=s$ in the equation $\wp(s)=0$, they are given by
\begin{equation}
 \label{E31}
x_{1,2}=-\frac{\Gamma^{32}_{0}+i\Delta_{p}}{2}\pm \sqrt{\Big[\frac{\Gamma^{32}_{0}-i\Delta_{p}}{2}\Big]^{2}-\frac{\Omega_{pmp}^{2}}{4}}.
\end{equation}
If the classical pump laser field is resonance with the atomic transition frequency; $\Delta_{p}=0$, we have
\begin{equation}
 \label{E32}
x_{1,2}=-\frac{\Gamma^{32}_{0}}{2}\pm \sqrt{\frac{(\Gamma^{32}_{0})^{2}}{4}-\frac{\Omega_{pmp}^{2}}{4}}
\end{equation}
Using the residue theorem, Eqs (\ref{E27}) and (\ref{E28}) are easily inverted to give
\begin{equation}
\label{E33}
 C_{2}(t)=\sum_{i} e^{x_{i} t}\frac{(x_{i}+i\Delta_{p})C_{2}(0)+i\frac{\Omega_{pmp}}{2}e^{-i\vartheta_{pmp}}C_{1}(0)}{(x_{i}-x_{j})},~~~ i,j=1,2;~~~ i\neq j
\end{equation}
\begin{equation}
\label{E34}
e^{-i\Delta_{p} t} C_{1}(t)=\sum_{i} e^{x_{i} t}\frac{[x_{i}+\Gamma^{32}_{0}]C_{1}(0)+i\frac{\Omega_{pmp}}{2}e^{i\vartheta_{pmp}}C_{2}(0)}{(x_{i}-x_{j})}~~~ i,j=1,2;~~~ i\neq j
\end{equation}
with
\begin{equation}
\label{E35}
 C_{1}(0)=\frac{1}{\sqrt{2}}e^{i\varphi}, ~~~~~~~~~C_{2}(0)=C_{1}(0) e^{-i\varphi}
\end{equation}
\par From equation (\ref{E32}) we see that both roots $x_{i}$ are (a) negative when $\Omega_{pmp}\leqslant\Gamma^{32}_{0}$ , and (b) complex (with a negative real part equal to $-\Gamma^{32}_{0}/2$) when $\Omega_{pmp}>\Gamma^{32}_{0}$. Thus the time evolution of amplitudes $C_{m}(t), m=1,2$ (and hence of the upper-level populations can be divided into two regimes of different behavior. For $\Omega_{pmp}>\Gamma^{32}_{0}$, the populations display pronounced oscillations before decaying to zero. On the other hand, when
$\Omega_{pmp}<\Gamma^{32}_{0}$ the populations barely complete an oscillation before decaying to zero \cite{MesfiSajeev03}. Thus, the driving field induces oscillations on the populations of the upper levels. The stronger the driving field (i.e. the larger the ), the faster the oscillations.
\subsubsection{Driving field with intensity equals to the vacuum spontaneous emission rate}
Thus, as we seek for comparison, we consider only the case when the pump laser field is applied with intensity; $\Omega_{pmp}$, equals to  the free space spontaneous emission rate $\Gamma^{32}_{0}$, i.e.,  $\Omega_{pmp}=\Gamma^{32}_{0}$. When this occurs, equation (\ref{E32}) has a double root $x_{1}=x_{2}=-\Gamma^{32}_{0}/2$ and
inversion of equations (\ref{E33}) and (\ref{E34}) gives
\begin{equation}
\label{E36}
C_{2}(t)=e^{-\frac{\Gamma^{32}_{0}}{2} t}\Big[C_{2}(0)-\frac{\Gamma^{32}_{0} t}{2}\Big(C_{2}(0)-i e^{-i\vartheta_{pmp}}C_{1}(0)\Big)\Big]
\end{equation}
\begin{equation}
\label{E37}
C_{1}(t)=e^{-\frac{\Gamma^{32}_{0}}{2} t}\Big[C_{1}(0)+\frac{\Gamma^{32}_{0} t}{2}\Big(C_{1}(0)+i e^{i\vartheta_{pmp}}C_{2}(0)\Big)\Big],
\end{equation}
which in agreement with results of Ref. \cite{MesfiSajeev03}.
\subsection{Three-level atom in front of a perfectly conducting planar dielectric half-space}
The presence of linear media in the form of macroscopic bodies changes the structure of the electromagnetic field compared to that in the free space and in consequence the electromagnetic vacuum felt by an atom is changed. As we mentioned above, absorption is always associated with additional noise which its effect included in the corresponding dynamical variables of the medium. In addition, the photonic density of states can be modified by the presence of macroscopic bodies, accordingly the mathematical treatment of the  integral kernel given by Eq. (\ref{E21}) will be completely modified, and as a consequence its physical interpretation. Starting from Eq. (\ref{E20}), on taking the time integral of both sides of this equation, it is not difficult to  see that the equation can be converted into the integral equation
\begin{equation}
\label{E38}
C_{2}(t)=C_{2}(0)+\frac{i\Omega_{pmp}}{2}e^{-i\vartheta_{pmp}}\int_{0}^{t} dt^{\prime}~e^{-i\Delta_{p}t^{\prime}}C_{1}(t^{\prime})
+\int_{0}^{t}~dt^{\prime}~\mathcal{K}(t,t^{\prime})C_{2}(t^{\prime})
\end{equation}
where
\begin{equation}
\label{E39}
\mathcal{K}(t,t^{\prime})= \frac{|d_{23}|^{2}}{\hbar\varepsilon_{0}\pi}\int_{0}^{\infty} d\omega\frac{\omega^{2}}{c^{2}}\frac{\mathrm{Im}~ \bm{G}(z_{A},z_{A},\omega)}{i(\omega-\omega_{0})}\biggl(e^{-i(\omega-\omega_{0}) (t-t^{\prime})}-1\biggr)
\end{equation}
From the above it is seen that $C_2 (t)$, with the presence of macroscopic bodies being fully included in
the Green tensor of the system, depends crucially on the actual structure of the Green tensor which carry all information about the macroscopic bodies properties. In general, it seems not easy to calculate $C_2 (t)$ in an exact analytical form. However, using limiting cases of Eq. (\ref{E39}) will simplify the problem to obtain the analytical solution without explicitly making use of the actual structure of the Green tensor.\\
In order to study the case where the atom is surrounded by matter, the atom should be assumed to be localized in a small free region, so that the Green tensor at the position of the atom reads
\begin{equation}
\label{E40}
 \bm{G}(z_{A},z_{A},\omega)=G^{V}(z_{A},z_{A},\omega)+G^{R}(z_{A},z_{A},\omega)
\end{equation}
where $G^{R}(z_{A},z_{A},\omega)$ describes the effect of reflection at the (surface of discontiniouty of the surrounding medium), and $G^{V}(z_{A},z_{A},\omega)$ is the vacuum Green tensor given by Eq.(\ref{E22}).
The contribution of $G^{V}$ to $\mathcal{K}(t,t^{\prime})$, in Eq. (\ref{E39}) takes the form
\begin{equation}
\mathcal{K}(t,t^{\prime})=-\frac{|d_{23}|^{2}}{\hbar\varepsilon_{0}\pi}\int_{0}^{\infty} d\omega\frac{\omega^{2}}{c^{2}}\frac{\mathrm{Im}~G^{V}(z_{A},z_{A},\omega)}{i(\omega_{0}-\omega)}\biggl(e^{i(\omega_{0}-\omega) (t-t^{\prime})}-1\biggr)\nonumber
\end{equation}
\begin{equation}
\label{E41}
- \frac{|d_{23}|^{2}}{\hbar\varepsilon_{0}\pi}\int_{0}^{\infty} d\omega\frac{\omega^{2}}{c^{2}}\frac{\mathrm{Im}~G^{R}(z_{A},z_{A},\omega)}{i(\omega_{0}-\omega)}\biggl(e^{i(\omega_{0}-\omega) (t-t^{\prime})}-1\biggr).
\end{equation}
This equation can be regarded as the basic equation for studying the influence of an arbitrary configuration of dispersing and absorbing matter on the spontaneous emission decay and related phenomena.
\subsubsection*{B.1. Weak atom-field coupling}
To achieve our goal, and hence, demonstrate the usefulness of this article simply, let us restrict our attention to the case when atom-field system is coupled weakly. This can be done mathematically by applying the Markov approximation. When this approximation applies, i.e., when in a coarse-grained description of the atomic motion memory effects are disregarded, then we may let
\begin{equation}
\label{E42}
 \frac{e^{i(\omega_{0}-\omega) (t-t^{\prime})}-1}{i(\omega_{0}-\omega)}\rightarrow \zeta(\omega_{0}-\omega)
\end{equation}
in Eq.(\ref{E42}) $[\zeta(x)=\pi\delta(x)+i\mathcal{P}\frac{1}{x}]$, and thus
\begin{equation}
\label{E43}
 \mathcal{K}(t,t^{\prime})=-\frac{\Gamma^{32}}{2}+i\delta\omega_{0}
\end{equation}
where $\Gamma^{32}$ and $\delta\omega_{0}$ are respectively given by
\begin{equation}
\label{E44}
 \Gamma^{32}=\Gamma^{32}_{0}+\frac{2|d_{23}|^{2}}{\hbar\varepsilon_{0}}\frac{\omega_{0}^{2}}{c^{2}}\mathrm{Im}~G^{R}(z_{A},z_{A},\omega_{0})
\end{equation}
\begin{equation}
\label{E45}
 \delta\omega_{0}=\frac{|d_{23}|^{2}}{\hbar\varepsilon_{0}\pi}\mathcal{P}\int_{0}^{\infty} d\omega\frac{\omega^{2}}{c^{2}}
\frac{\mathrm{Im}~G(z_{A},z_{A},\omega)}{(\omega-\omega_{0})}
\end{equation}
Obviousely, in contrast to free space, the decay rate now becomes a function of the atomic position and the orientation of the transition dipole moment. In other words, the vacuum fluctuations felt by an atom in the strictly free space are inhomogeneously and anisotropically changed by the presence of the bodies, in general. The nearer to a body that the atom is located, the stronger the effect to be expected.
\par Following the approach of Dung {et al.,} \cite{DKW01}, and  after Recalling the Kramers-Kronig relation for the Green tensor, we may approximately rewrite $\delta\omega_{0}$ as
\begin{equation}
\label{E46}
\delta\omega_{0}=\frac{|d_{23}|^{2}\omega_{0}^{2}}{\hbar\varepsilon_{0} c^{2}}\mathrm{Re}~G^{R}(z_{A},z_{A},\omega_{0})
-\frac{|d_{23}|^{2}}{\pi}\int_{0}^{\infty} d\omega\frac{\omega^{2}}{\omega_{0}^{2}}
\frac{\mathrm{Im}~G^{R}(z_{A},z_{A},\omega)}{(\omega+\omega_{0})}.
\end{equation}
The second term, which is only weakly sensitive to the atomic transition frequency, is small compared to the first one and can therefore  be generally neglected. Obviously, the divergent contribution of vacuum to the line shift is included in the atomic transition frequency $\omega_{0}$. Subistituting (\ref{E43}) into (\ref{E38}) and taking the time derivative of both sides with the setting of $\Gamma^{32}/2= \varpi^{32}$ for simplicity, we obtain
\begin{equation}
\label{E47}
\dot{C}_{2}(t)=i\frac{\Omega_{pmp}(t)}{2} e^{-i\vartheta_{pmp}}~e^{-i\Delta_{p}t}C_{1}(t)
+(-\varpi^{32}+i\delta\omega_{0})~C_{2}(t),
\end{equation}
the next step is to apply the Laplace transforms on Eqs. (\ref{E14}) and (\ref{E47}), with the same initial conditions, to obtain
\begin{equation}
\label{E48}
\widetilde{C}_{2}(s)=\frac{(s+i\Delta_{p})C_{2}(0)+i\frac{\Omega_{pmp}}{2}e^{-i\vartheta_{pmp}}C_{1}(0)}{\wp(s)},
\end{equation}
and
\begin{equation}
\label{E49}
\widetilde{C}_{1}(s+i\Delta_{p})=\frac{[s-(-\varpi^{32}+i\delta\omega_{0}]C_{1}(0)+i\frac{\Omega_{pmp}}{2}e^{i\vartheta_{pmp}}C_{2}(0)}{\wp(s)},
\end{equation}
where
\begin{equation}
\label{E50}
\wp(s)=s^2+\Big[(\varpi^{32}-i\delta\omega_{0})+i\Delta_{p}\Big]s+\Big[i\Delta_{p}(\varpi^{32}-i\delta\omega_{0})+\frac{\Omega_{pmp}^{2}}{4}\Big]=\prod_{k=1}^{2}(s-y_{k}),
\end{equation}
with $y_{k}$ are given by
\begin{equation}
\label{E51}
y_{1,2}=-\frac{(\varpi^{32}-i\delta\omega_{0})+i\Delta_{p}}{2}\pm \sqrt{\Big[\frac{(\varpi^{32}-i\delta\omega_{0})-i\Delta_{p}}{2}\Big]^{2}-\frac{\Omega_{pmp}^{2}}{4}}.
\end{equation}
Following the same procedure, as previous, it is not difficult to obtain $C_{m}(t), m=1,2$ in the form
\begin{equation}
\label{E52}
 C_{2}(t)=\sum_{i} e^{y_{i} t}\frac{(y_{i}+i\Delta_{p})C_{2}(0)+i\frac{\Omega_{pmp}}{2}e^{-i\vartheta_{pmp}}C_{1}(0)}{(y_{i}-y_{j})},~~~i,j=1,2~~~i\neq j,
\end{equation}
\begin{equation}
\label{E53}
e^{-i\Delta_{p} t} C_{1}(t)=\sum_{i} e^{y_{i} t}\frac{[y_{i}-(-\varpi^{32}+i\delta\omega_{0})]C_{1}(0)+i\frac{\Omega_{pmp}}{2}e^{i\vartheta_{pmp}}C_{2}(0)}{(y_{i}-y_{j})},~~~i,j=1,2~~~i\neq j.
\end{equation}
\subsubsection*{B.1.1. Model permittivity of Drude-Lorentz type}
To be more specific, let the classically-pumped three-level atom located near the surface of an infinite half-space dielectric medium such that
\begin{equation}
\label{E54}
\varepsilon({\bf r},\omega)=
 \begin{cases}
 \varepsilon(\omega)&\text{if} ~~z\leq 0\cr
 1 &\text{if} ~~z> 0\cr
 \end{cases}
\end{equation}
To give an impression of what can be observed in real situation, let choose a planar dielectric surface of permittivity modeled by the widely-used-in-practice Lorentz type as
\begin{equation}
\label{E55}
 \varepsilon(\omega)=1+\frac{\omega_{p}^{2}}{\omega_{T}^{2}-\omega^{2}-i\omega\gamma_{}}
\end{equation}
\vspace{0.5pt}
where $\omega_{T}$ and $\gamma_{}$ are the medium oscillation frequencies and linewidths, respectively, and $\omega_{p}$ correspond to the coupling constants. An example of the permittivity for a (single resonance) dielectric as a function of frequency is shown in Fig. 3.
\vspace{-0.3cm}
\begin{figure}[tpbh]
\noindent
\begin{center}
\includegraphics[width=0.33\linewidth]
{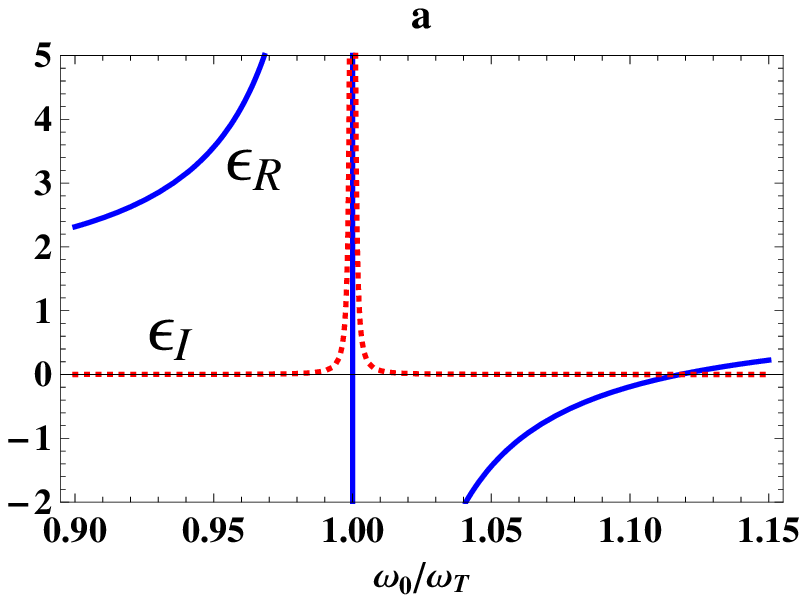}
\hspace{1.0cm}
\includegraphics[width=0.33\linewidth]
{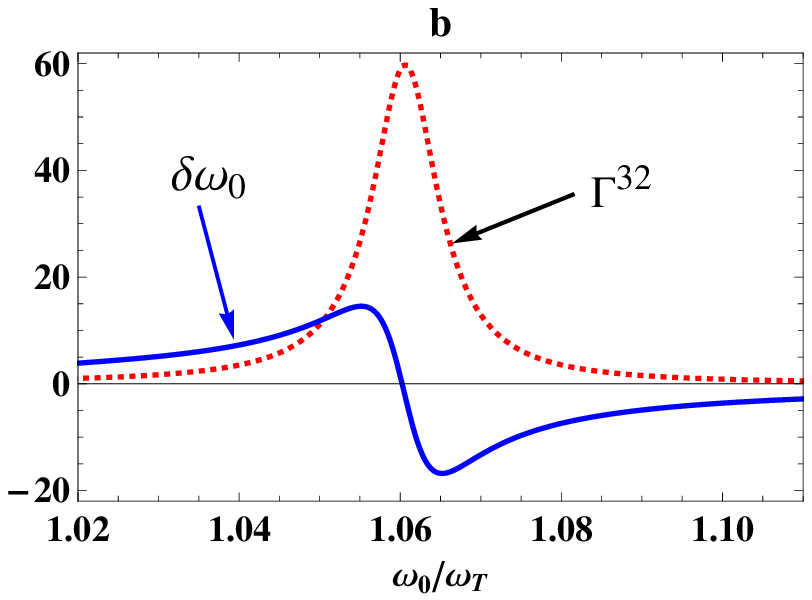}
\end{center}
\vspace{-0.7cm}
\caption{(a) Real and imaginary parts of the permittivity of single-resonance Drude-Lorentz type dielectric for $\omega_{p}=0.5\omega_{T}$, $\gamma=10^{-4}\omega_{T}$. The band gap covers the interval from $\omega_{L}=\omega_{T}$ to $\omega_{L}\simeq1.12 \omega_{T}$, (b) Spontaneous decay rate $\Gamma^{32}$ [Eq. (\ref{E56})] and level shift $\delta\omega_{0}$ [Eq. (\ref{E57})] for same $\omega_{p}$ while $\gamma=10^{-2}\omega_{T}$ and $z_{A}=0.05 \lambda_{T}$.}
\label{Fig.2}
\end{figure}
For $z>0$, and when the distance of the atom from the surface is small compared with the wave length, $kz\ll 1$, using the results obtained by Scheel {\it et al.} \cite{SKW99c} of the reflection part of the Green tensor yields for the decay rate 
\begin{equation}
\label{E56}
 \Gamma^{32}\approx\frac{3\Gamma^{32}_{0}}{8}\biggl(1+\frac{d_{z}^{2}}{d^{2}}\biggr)\biggl(\frac{c}{\omega_{0} z}\biggr)^{3}\times \frac{\varepsilon_{I}(\omega_{0})}{|\varepsilon(\omega_{0})+1|^{2}}
\end{equation}
The leading term $(\approx z^{-3})$, which is proportional to $\varepsilon_{I}(\omega_{0})$, is closely related to the  virtual photon emission with subsequent medium quasiparticle excitation (nonradiative decay), i.e., energy transfer from the atom to the medium.
\\
Similarly, for the line shift, due to the presence of the macroscopic bodies reads
\begin{equation}
\label{E57}
 \delta\omega_{0}\approx\frac{3\Gamma^{32}_{0}}{32}\biggl(1+\frac{d_{z}^{2}}{d^{2}}\biggr)\biggl(\frac{c}{\omega_{0} z}\biggr)^{3}\times \frac{|\varepsilon(\omega_{0})|^{2}-1}{|\varepsilon(\omega_{0})+1|^{2}}
\end{equation}
\section{The geometric phase}
If the final wavefunction cannot be obtained from the initial wavefunction by a multiplication with a complex number, the initial and final states are distinct and the evolution is noncyclic. The total phase has both dynamical, $\phi_d$, and geometric phase, $\phi_g$, parts for an arbitrary quantum evolution of the system from an original state at $t=0$ to a final wavefunction at time $t$.  Suppose state $\mid\psi(0)\rangle$ evolves to a state $\mid\psi(t)\rangle$ after a certain time t. If the scalar product \cite{SjPaEkAnEr00}
\begin{equation}
\label{E58}
\langle\psi(0)\mid \exp\Big(\frac{i}{\hbar}\int_{0}^{t}\hat{H}(t{}')dt{}'\Big)\mid\psi(0)\rangle
\end{equation}
can be written as $\beta ~e^{i\eta}$, where $\beta$ is a real number, then the noncyclic phase due to the
evolution from $\mid\psi(0)\rangle$ to $\mid\psi(t)\rangle$ is the angle $\eta$.  This noncyclic phase generalizes the cyclic
geometric phase since the latter can be regarded as a special case of the former in which $\beta=1$. The total phase (the topological Berry phase) acquired during an arbitrary evolution of a wavefunction from the vector $\mid\psi(0)\rangle$ to $\mid\psi(t)\rangle$ is prescribed by Phancharatnam as
\begin{equation}
 \label{E59}
\phi_{t}=\arg\langle\psi(0)\mid\psi(0)\rangle
\end{equation}
On substraction the dynamical phase $\phi_d$ from the total phase $\phi_{t}$ we obtain the geometric phase $\phi_g$. An exact
expression of the geometric phase can be obtained as
\begin{equation}
 \phi_{t}=\arg\{C_{1}^{\ast}(0)C_{1}(t)+C_{2}^{\ast}(0)C_{2}(t) e^{-i\omega_{0} t}\}\nonumber
\end{equation}
\begin{equation}
\label{E60}
=-\arcsin\biggl(\frac{Y(t)}{\sqrt{X^{2}(t)+Y^{2}(t)}}\biggr),
\end{equation}
with
\begin{equation}
 \label{E61}
X(t)=\mathrm{Re}[C_{1}^{\ast}(0)C_{1}(t)+C_{2}^{\ast}(0)C_{2}(t)e^{-i\omega_{0} t}],
\end{equation}
and
\begin{equation}
 \label{E62}
Y(t)=\mathrm{Im}(C_{1}^{\ast}(0)C_{1}(t)+C_{2}^{\ast}(0)C_{2}(t)e^{-i\omega_{0} t}),
\end{equation}
where the dynamical phase for an arbitrary quantum evolution from a time $t=0$ to a time $t$ is given by the time integral of the expectation value of the Hamiltonian over the time interval $[0,t]$
\begin{equation}
 \label{E63}
\phi_{d}=-\int_{0}^{t}\langle\psi(t)\mid\hat{H}(t)\mid\psi(t)\rangle dt
\end{equation}
As might be expected, the behavior of the present three-level system changes dramatically depending on the variation of the adjustable parameters of the system, namely, dipole vector orientation, the atom position, $z_A$ from the surface, and finally, the surface permittivity.
\begin{figure}[tpbh]
\noindent
\begin{center}
\includegraphics[width=0.32\linewidth]
{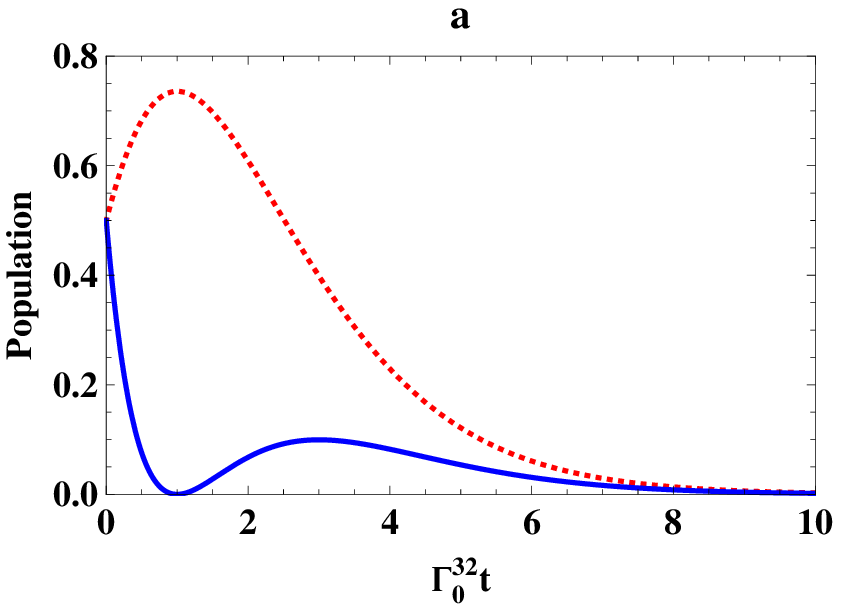}
\includegraphics[width=0.32\linewidth]
{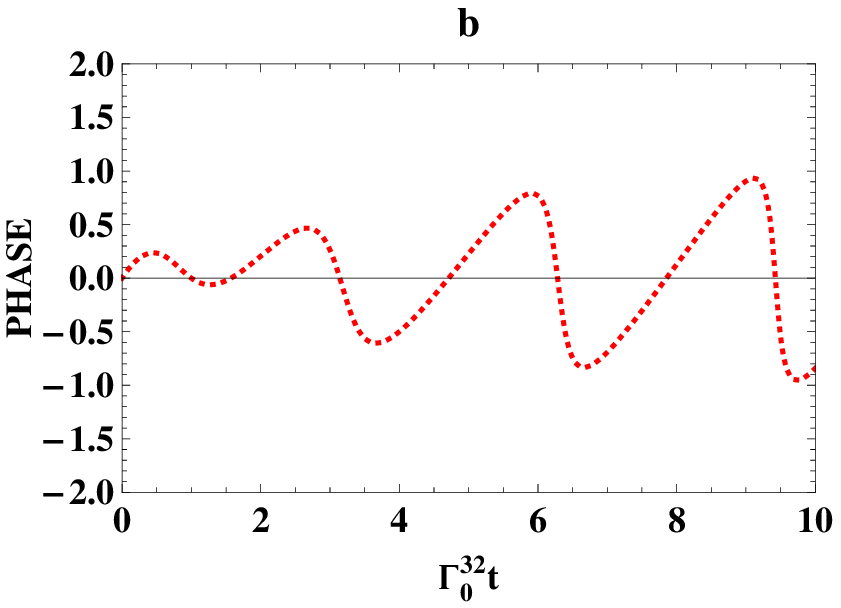}
\includegraphics[width=0.32\linewidth]
{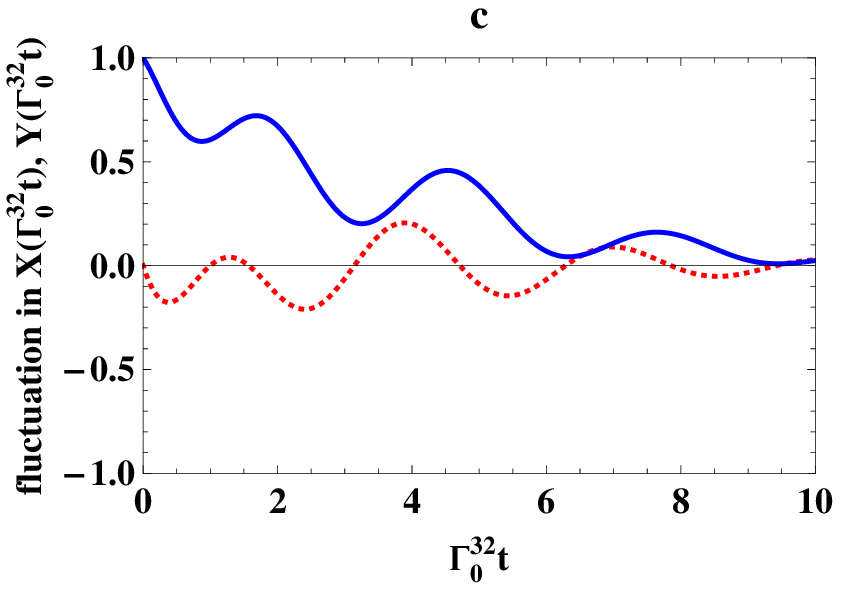}
\end{center}
\vspace{-0.7cm}
\caption{Two dimension plot of $\mid 2\rangle$ state population (a), the topological Berry phase $\phi_{t}$ (b) and the fluctuations in $X$ and $Y$ for atom in free space with $\Omega_{pmp}/\Gamma_{0}^{32}=1$.}
\label{Fig.3}
\end{figure}\\
Examples [for all our plots we have used fixed values for $C_{1}(0)=C_{2}(0)=\frac{1}{\sqrt{2}}$ and $\phi=\pi/2$] of the populations and the corresponding topological Berry phase against $\Gamma_{0}^{32} t$ are shown in Fig. 3 for an atom in free space. It is rather interesting to mention to the fact that: in the Feynman approach to quantum mechanics it is important not only the modulus and argument of complex number, but the quantum fluctuations of this variables too. In such a representation, the deviation of quantum trajectories from the classical trajectories are described by the quantum fluctuations \cite{Eleuch09} . Therefore we have further investigated the fluctuations of the real and imaginary parts of $\arg\langle\psi(t)\mid\psi(t)\rangle$. At first, for initial $C_{2}(0)=0$, the topological Berry phase $\phi_{t}$ vanishes, this means physically that there is no phase in photon transition that make the photon transition is in a straight line, i.e., the photon transition will occurs without any phase. Thus the atomic coherence is necessary for the development of the topological Berry phase $\phi_{t}$. Note the populations decay exponentially while $\phi_t$ normally oscillates, due to the oscillations in $X$ and $Y$, with an amplitude increases with time.
\par In an interesting proposal for detecting the topological Berry’s phase was suggested \cite{EkJo96,Fuboved00} for a certain adiabatic evolution of a joint state of the internal levels of a trapped ion and its vibrational motion. It has been shown that despite being the phase gained by a joint state, its value is fundamentally dependent on the harmonic oscillator nature of the vibrational mode. \par When the effect of dielectric matter is considered, we clearly notice very different figure shape. It worth to note that, in the short range interval of $\omega_{0}/\omega_{T}$ at which the decay rate makes half period and change its evolution direction, the level shift $\delta\omega_{0}\approx 0$, in this interval the complex amplitudes $C_{m}(t), m=1,2$, Eqs. (\ref{E52}, \ref{E53}), after putting $\Delta_{p}=0$, depend only on the decay rate $\Gamma^{32}$, and can take the same form of Eqs. (\ref{E36}, \ref{E37}) of free space   with the replacement of the free space decay rate $\Gamma^{32}_{0}$ by $\varpi^{32}$. 
\begin{figure}[tpbh]
\noindent
\begin{center}
\includegraphics[width=0.32\linewidth]
{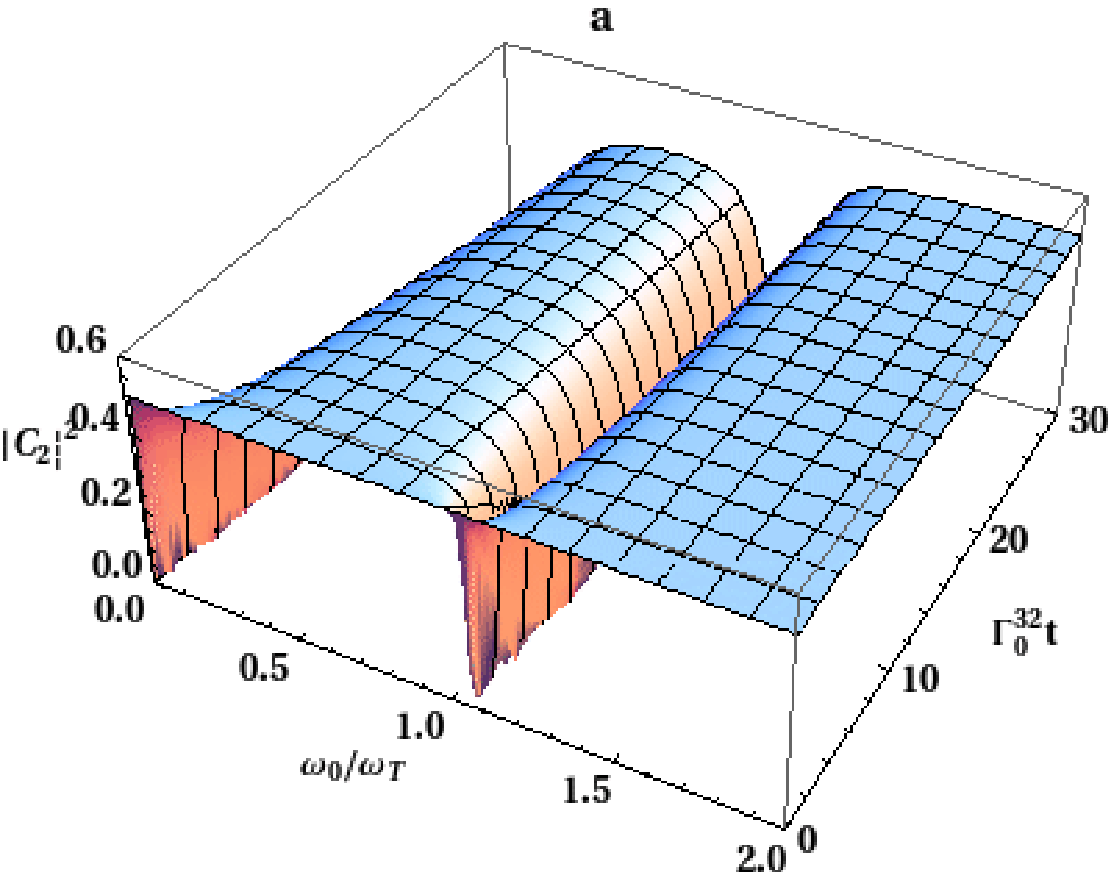}
\hspace{1.0cm}
\includegraphics[width=0.32\linewidth]
{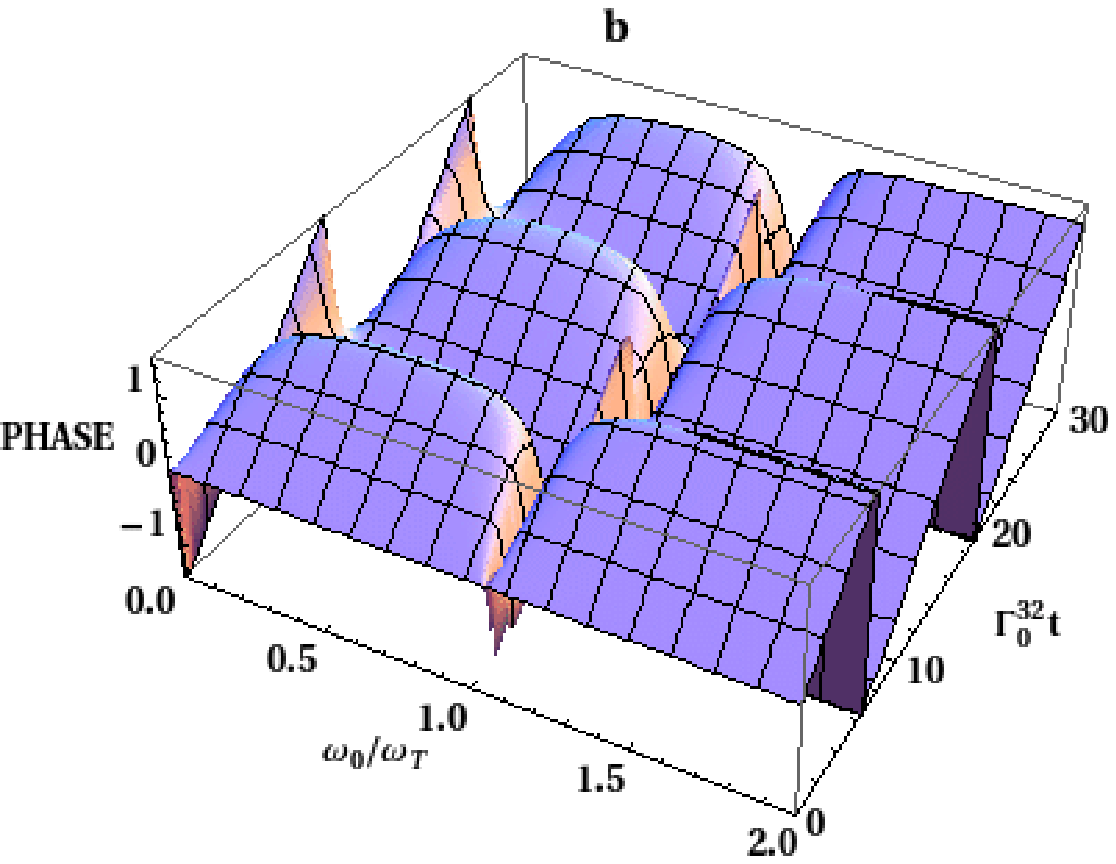}
\end{center}
\vspace{-0.5cm}
\caption{Mesh plot of $\mid 2\rangle$ state population (a) and the topological Berry phase $\phi_{t}$ (b) as functions of the transition frequency  $\omega_{0}/\omega_{T}$ and $\Gamma_{0}^{32} t$ near a planar dielectric half-space for an $x$-oriented transition dipole moment, with  $\Omega_{pmp}=\varpi^{32}$, $\omega_{p}=0.5\omega_{T}$, $z=0.05\lambda_{T}$ and  $\gamma=10^{-3}\omega_{T}$.}
\label{Fig.4}
\end{figure}
Examples of the population, $\phi_{t}$ and the fluctuations in $X$ and $Y$ for dielectric matter of Drude-Lorentz type, with linewidth $\gamma=10^{-3}$, are shown as mesh plot in Figs. 4, 6 and 8 with the corresponding cross section (the dependance on the transition frequency) when $\Gamma_{0}^{32} t=10.0$ in Figs. 5, 7 and 9. The property of Drude-Lorentze type, where it features band gab between the transverse frequency $\omega_{T}$ and longitudinal frequency $\omega_{L}=\sqrt{\omega_{T}^{2}+\omega_{p}^{2}}$, where $\mathrm{Re}~ \varepsilon(\omega)<0$, control strongly the behavior of the functions under study. 
\begin{figure}[tpbh]
\noindent
\begin{center}
\includegraphics[width=0.32\linewidth]
{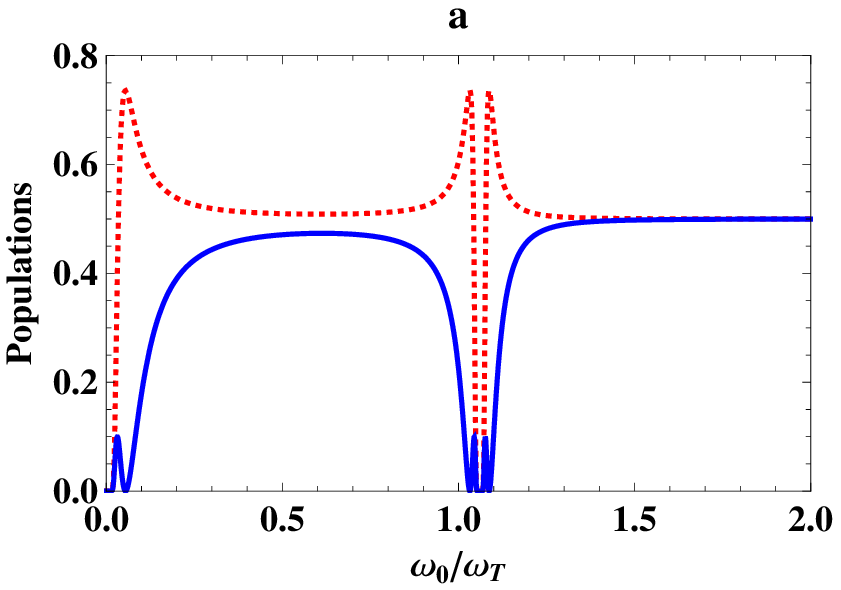}
\includegraphics[width=0.32\linewidth]
{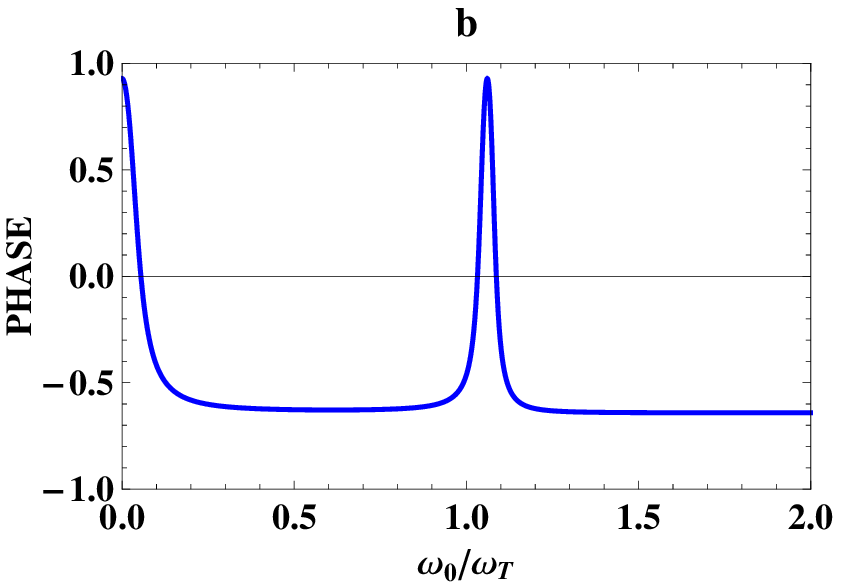}
\includegraphics[width=0.32\linewidth]
{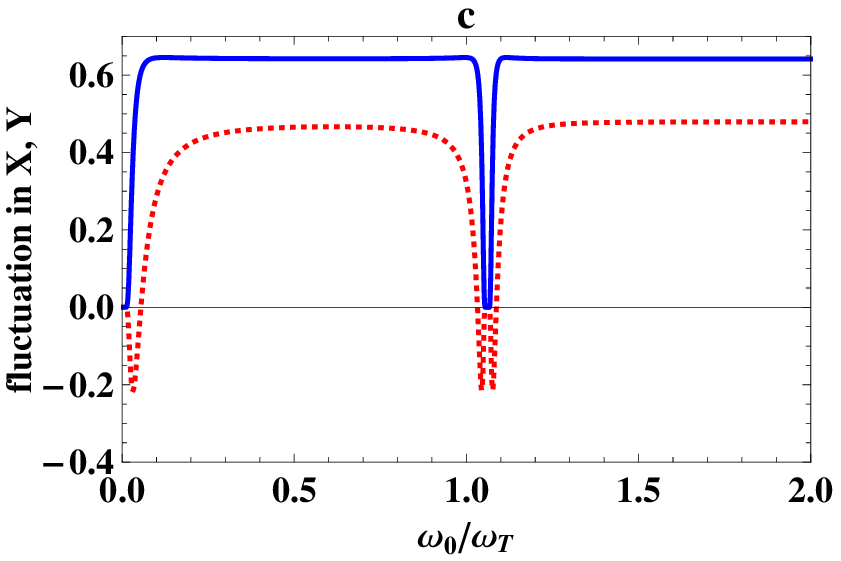}
\end{center}
\vspace{-0.7cm}
\caption{Cross section of Fig. 4 as well as fluctuations in $X$ and $Y$  (c), when $\Gamma_{0}^{32} t=10.0$ while rest of parameters as in Fig. 4.}
\label{Fig.5}
\end{figure}
Moreover, since the Drude-Lorentze model incorporates surface-guided waves, which are observed in the presence of the interface and bound to it, with the amplitudes damped into either of the neighboring media, we notice the populations, $\phi_{t}$ and the fluctuations in $X$ and $Y$ affected strongly with these properties and appear as surface-guided functions. It is seen that, in the band-gap region, due to the drastically increase if of the spontaneous decay rate, because of the non-radiative decay channel associated with absorption, the population amplitude damped into the band-gap region while $\phi_{t}$ approaches unity as  seen from the figures. 
Fig. 5 also shows that the populations vanishes inside the band gap in the interval where $\omega_{L}\in[\omega_{T},1.12 \omega_{T}]$,  as expected, while fluctuations go beyond negative values.
\begin{figure}[tpbh]
\noindent
\begin{center}
\includegraphics[width=0.32\linewidth]
{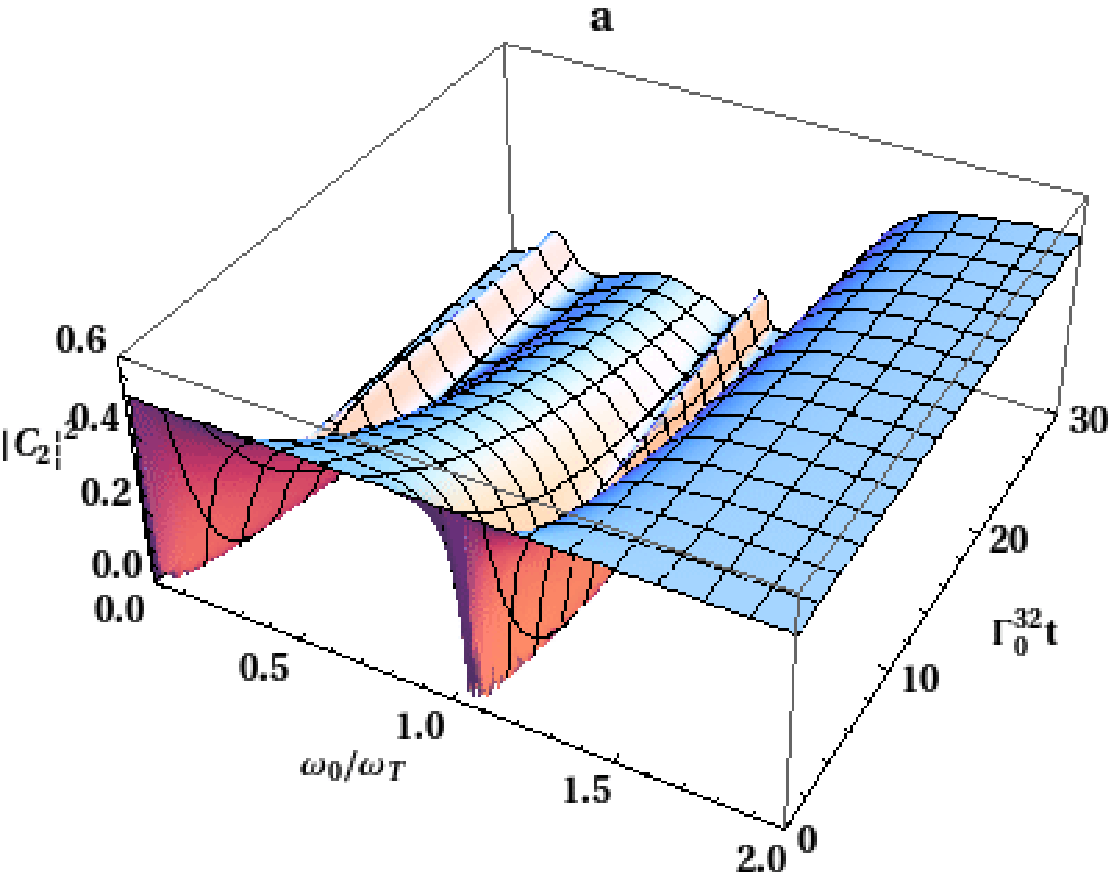}
\hspace{1.0cm}
\includegraphics[width=0.32\linewidth]
{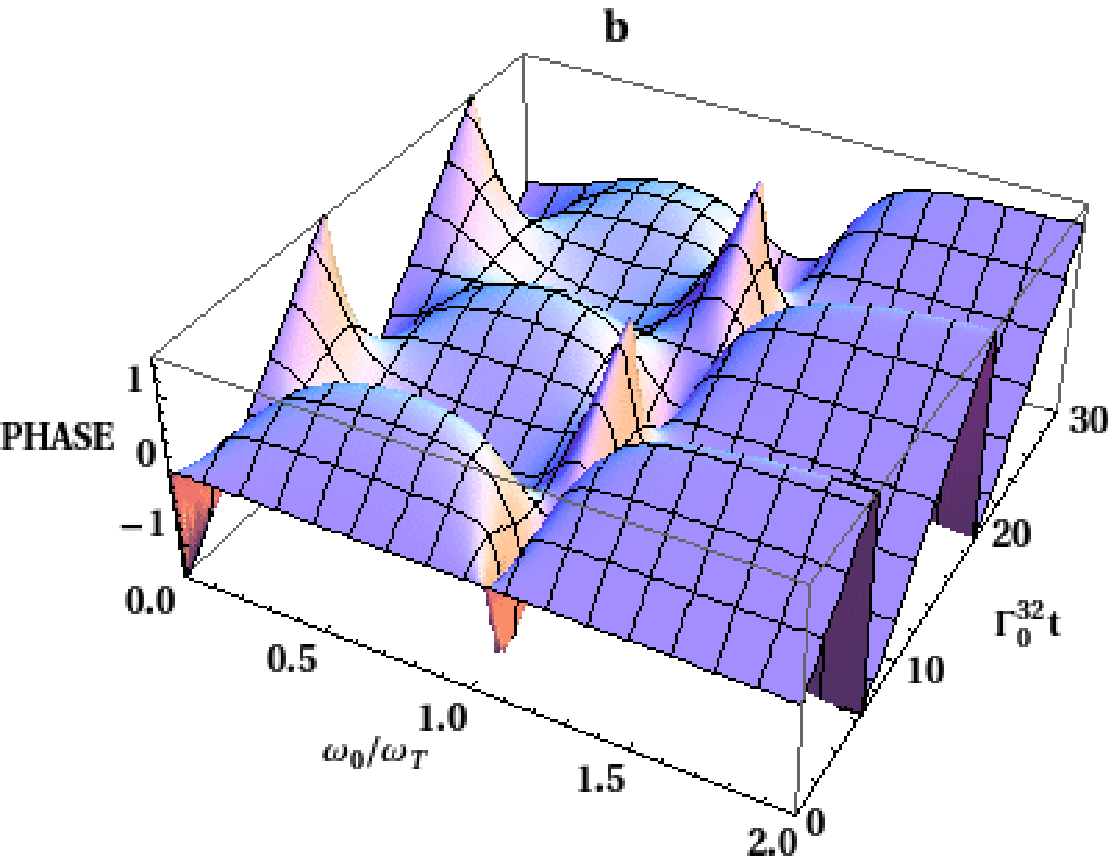}
\end{center}
\vspace{-0.5cm}
\caption{The same as Fig. 4 but for $\gamma=10^{-2}\omega_{T}$}
\label{Fig.6}
\end{figure}
On increasing the damping parameter,$\gamma$, one observes, the narrow-band gradually getting so much broaden [Figs. 7; $\gamma=10^{-2}$  and 9; $\gamma=10^{-1}$ ] that it eventually disappears (not shown), due to the increase of spontaneous decay for frequency $\omega_{0}$ - with value equals to unity  for $\phi_{t}$ is preserved- where most of the energy emitted by the atom is absorbed by the medium wall in the course of time while a fraction of light can scape to free space which reflects the clearly-noticed decay of the population as time passes [see Figs. 6(a) and 8(a)]. Moreover, in the absorption band and in the vicinity of the absorption band, i.e., in the region between resonance $\omega_{T}$ and longitudinal frequency $\omega_{L}=\sqrt{\omega_{T}^{2}+\omega_{p}^{2}}$ (in the figures, $\omega_{L}=1.12\omega_{T}$), these functions respond sensitively to the change of the damping parameter $\gamma$. The differences between the cases here are less pronounced for strong absorption, i.e., when the value of the bandwidth parameter $\gamma$ is sufficiently large (compare Figs. 5, 7 and 9). We conclude that the differences obtained differ essentially in the way the bandwidth parameter is introduced.
\begin{figure}[tpbh]
\noindent
\begin{center}
\includegraphics[width=0.32\linewidth]
{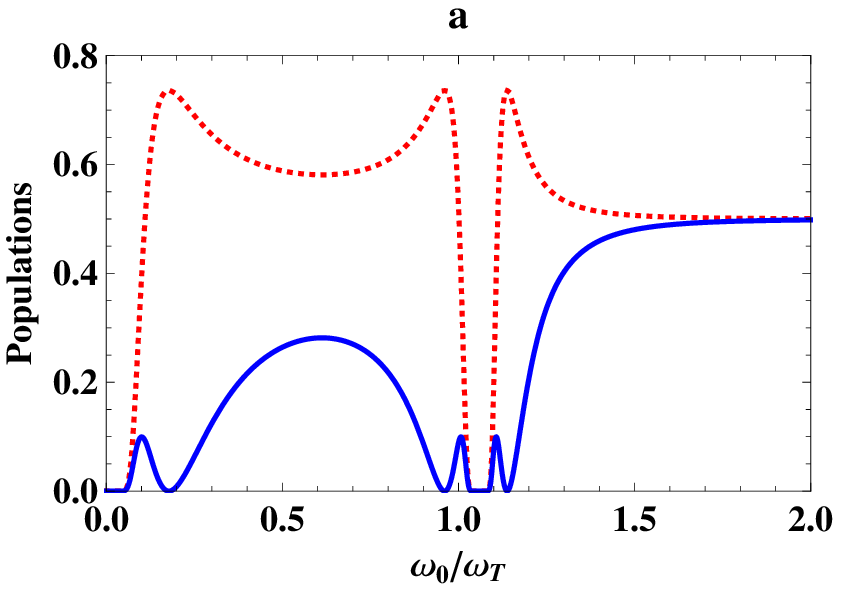}
\includegraphics[width=0.32\linewidth]
{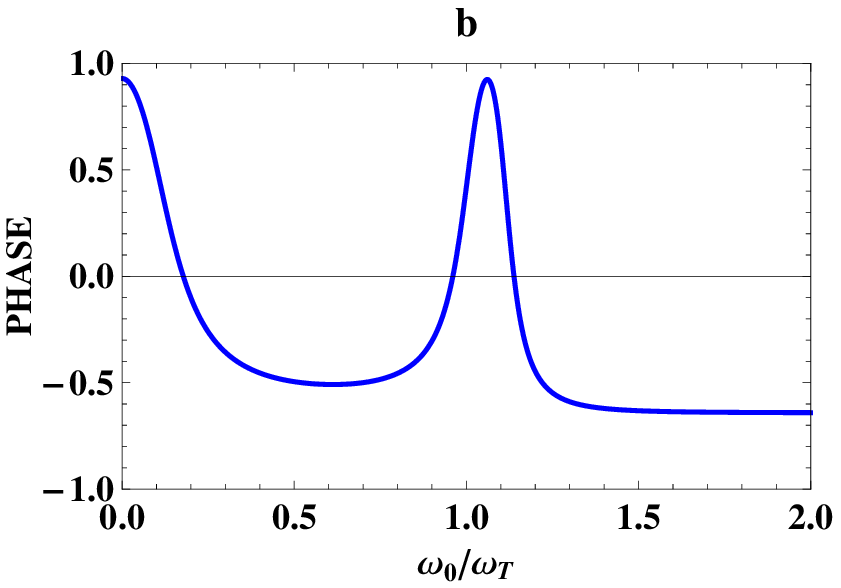}
\includegraphics[width=0.32\linewidth]
{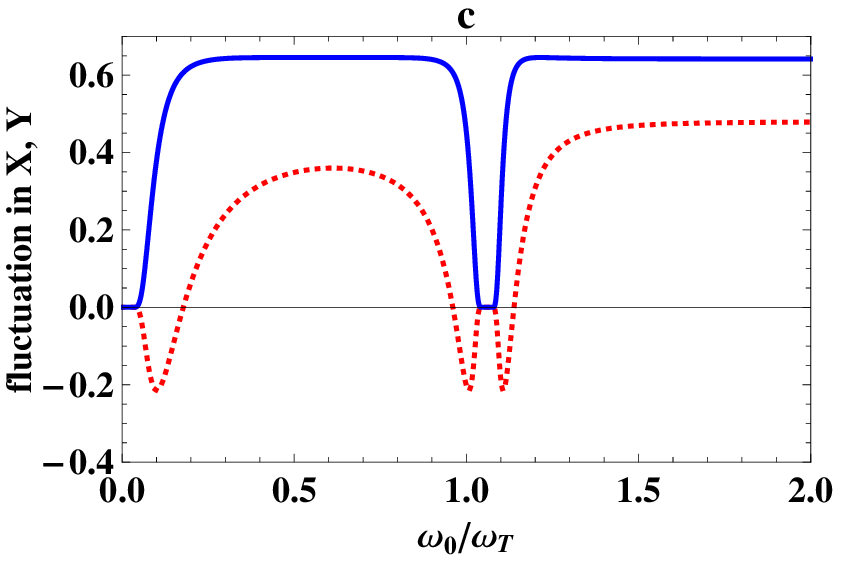}
\end{center}
\vspace{-0.7cm}
\caption{The same as Fig. 5 but for $\gamma=10^{-2}\omega_{T}$}
\label{Fig.7}
\end{figure}
The maxima of the populations  below the band gap and at the field resonance ($\omega_{0}\approx\omega_{T}$) frequencies (Figs. 4 and 5) indicate that, although the decay can be noticeably enhanced, the probability of emission of a really observable photon can be substantially reduced compared to the case of spontaneous emission in the free space. Obviously, a photon emitted at such a frequency is typically captured by the surface for some time due to the deep  penetration of radiation into the surface, and hence the probability of photon absorption is increased. Also, for transition frequencies inside the band gap, two regions can be distinguished. 
\begin{figure}[tpbh]
\noindent
\begin{center}
\includegraphics[width=0.32\linewidth]
{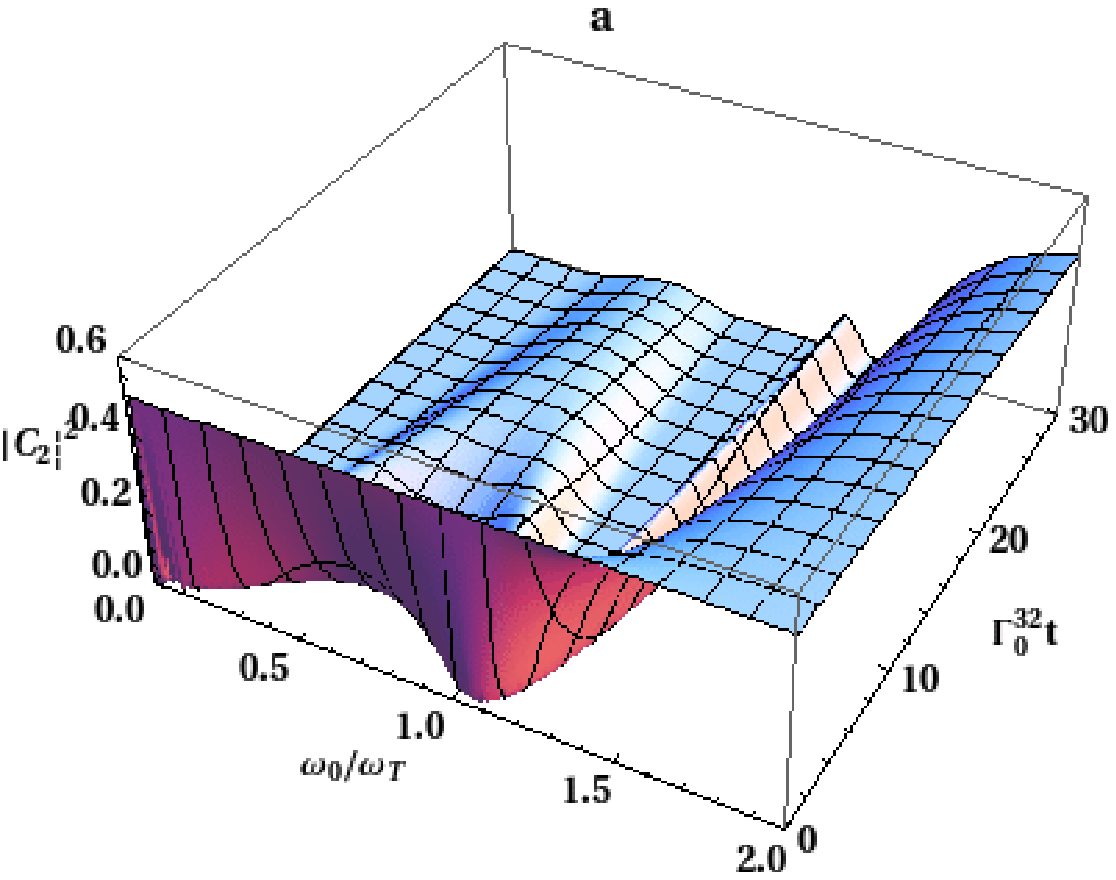}
\hspace{1.0cm}
\includegraphics[width=0.32\linewidth]
{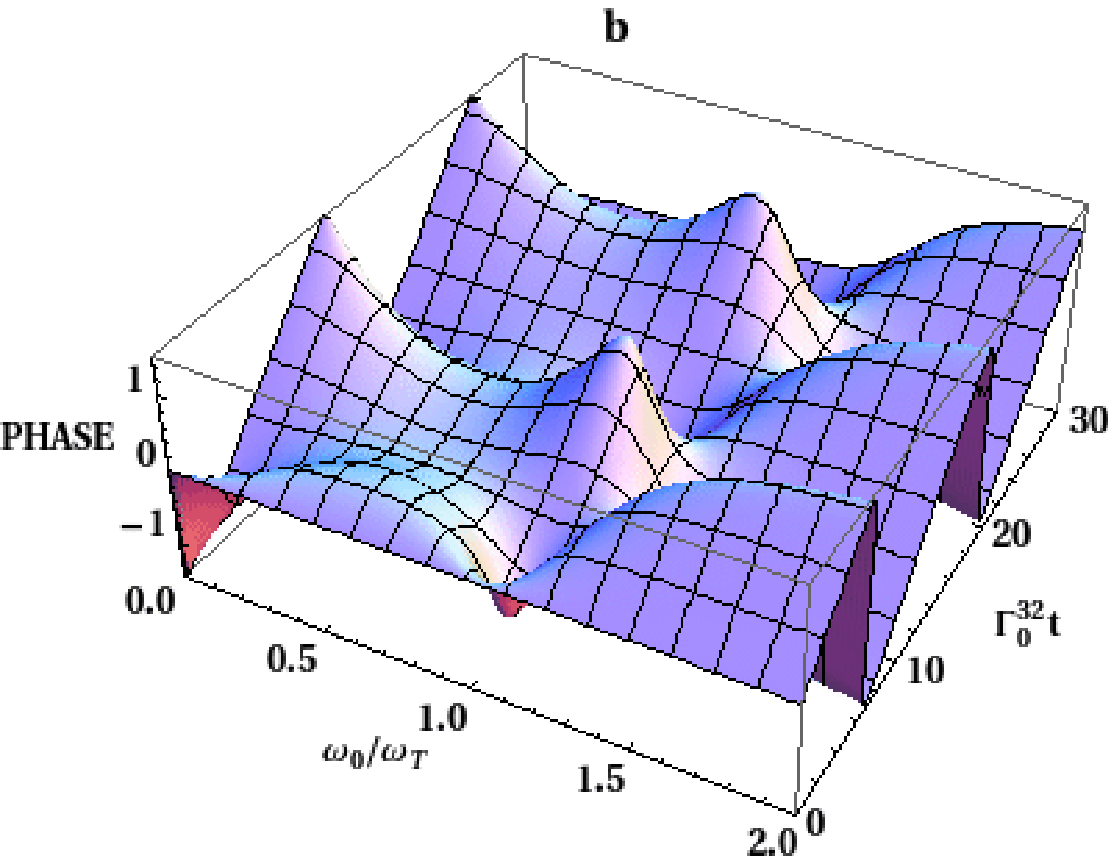}
\end{center}
\vspace{-0.5cm}
\caption{The same as Fig. 4 but for $\gamma=10^{-1}\omega_{T}$}
\label{Fig.8}
\end{figure}
In the low-frequency region ($\omega_{0}<\omega_{T}$), where surface-guided waves are typically excited, radiative decay dominates, i. e., the atomic transition is accompanied by the emission of a photon escaping from the system. Here, the radiation penetration depth into the surface is small and the probability of a photon being absorbed is also small. With increasing of atomic transition frequency the penetration depth increases and the chance of a photon to escape drastically diminishes. 
\begin{figure}[tpbh]
\noindent
\begin{center}
\includegraphics[width=0.32\linewidth]
{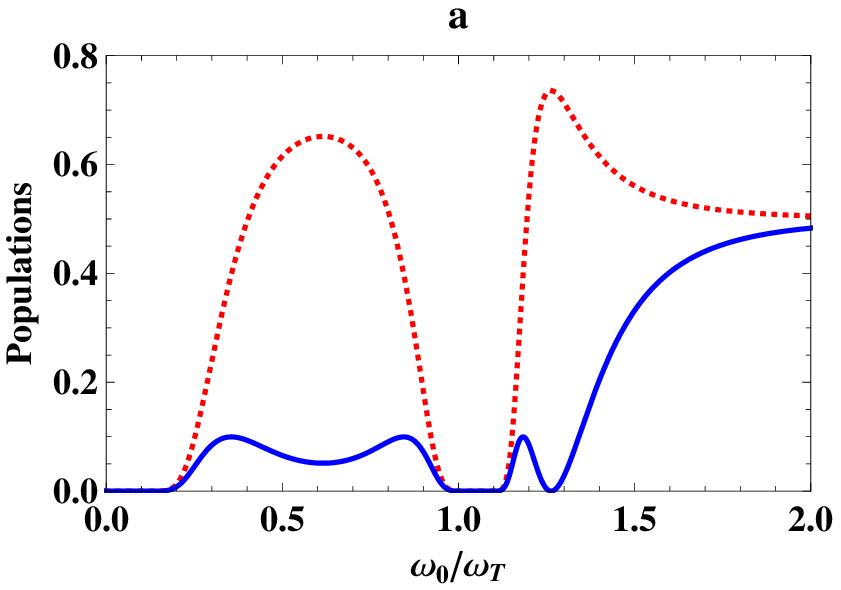}
\includegraphics[width=0.32\linewidth]
{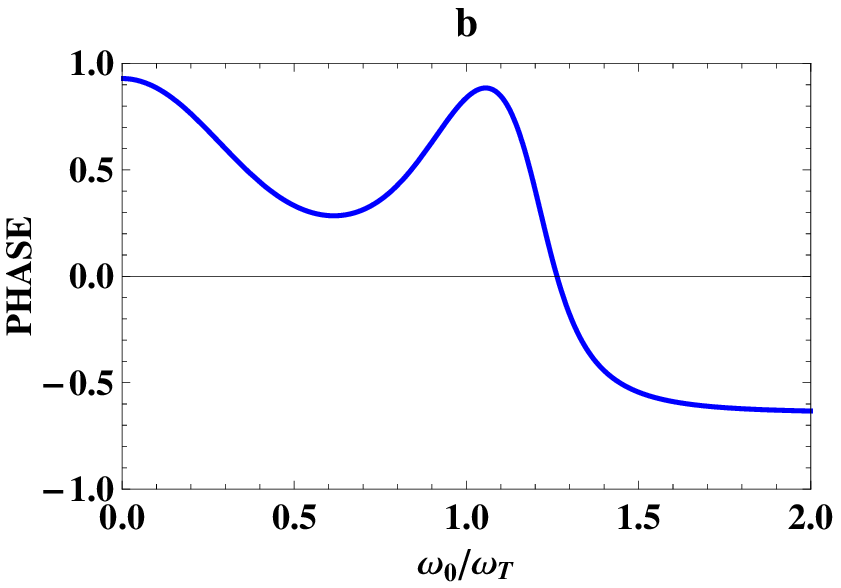}
\includegraphics[width=0.32\linewidth]
{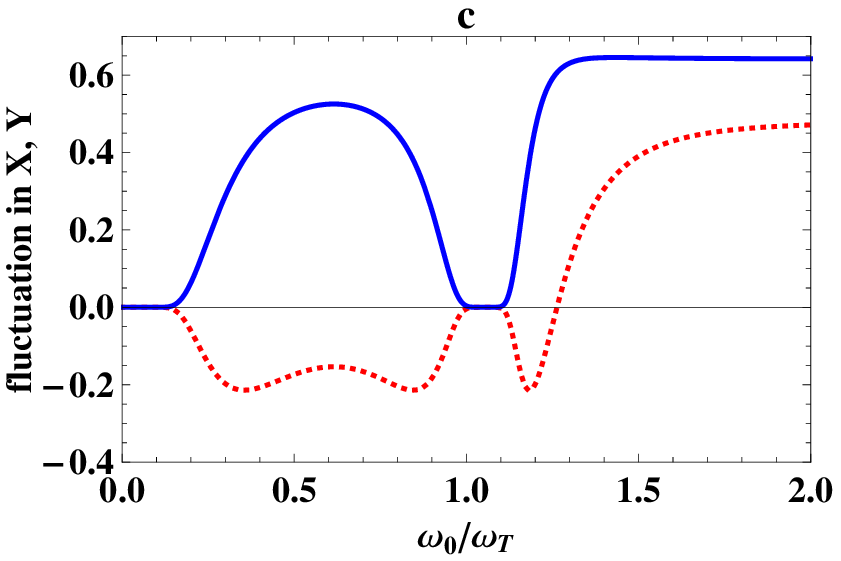}
\end{center}
\vspace{-0.7cm}
\caption{The same as Fig. 5 but for $\gamma=10^{-1}\omega_{T}$}
\label{Fig.9}
\end{figure}
As a result, photon absorption dominates \cite{HKW01}. In this case, in the regions below and above the band gap,  $\phi_{t}$ preserves a fixed negative short angle with $|\phi_{t}|\approx 77^{o}$,  which means that the wavefunction evolution and the original state move apart from each other, while inside the band gap region this angle decreases to reach a short angle equals to $\phi_{t}\approx 52^{o}$ (see Fig. 5). On increasing $\gamma$, $\phi_{t}$ differs considerably in the two regions, where below band gap $\phi_{t}$ shiftes to become smaller than above band gap region (Figs. 5 and 9) due to the strong penetration of the emitted photon into the surface.  We can say that the topological Berry phase strongly depends on the presence of dielectric surface and such systems are potentially interesting for their ability to process information in a novel way and might find application in models of quantum logic gates.
\begin{figure}[tpbh]
\noindent
\begin{center}
\includegraphics[width=0.32\linewidth]
{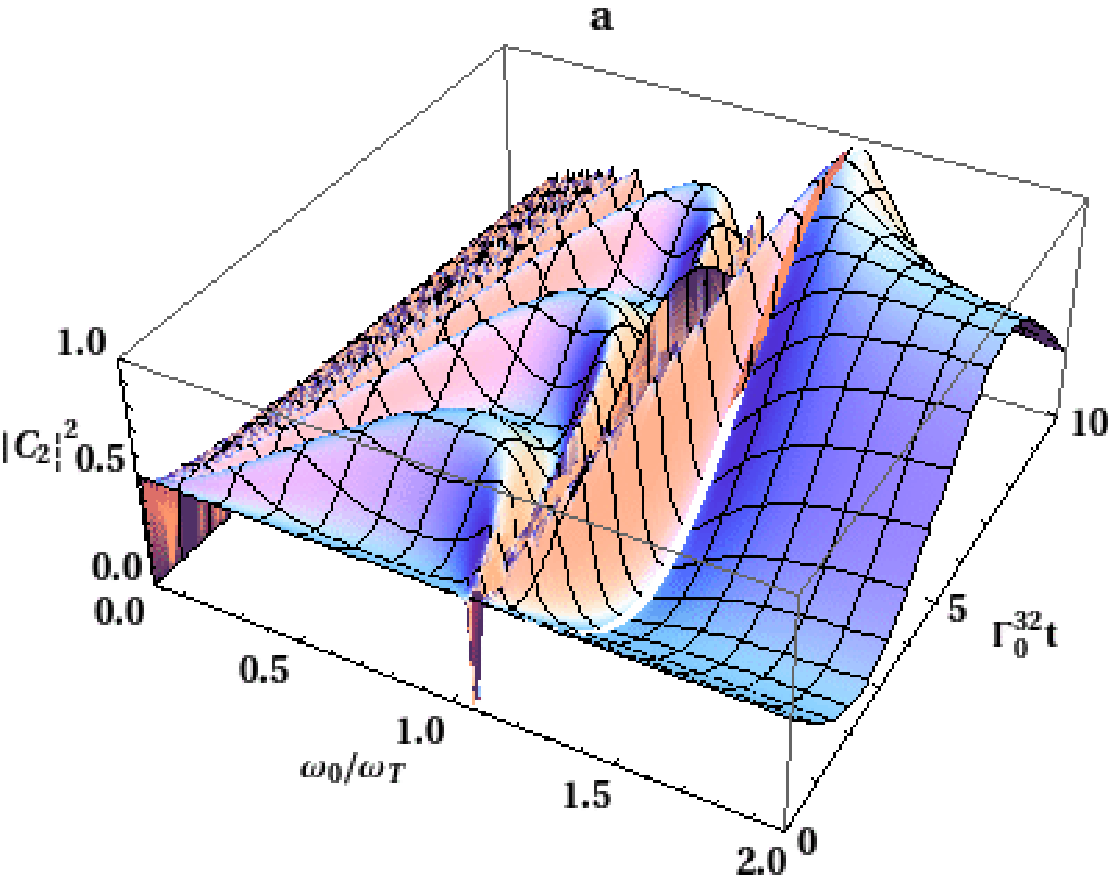}
\hspace{1.0cm}
\includegraphics[width=0.32\linewidth]
{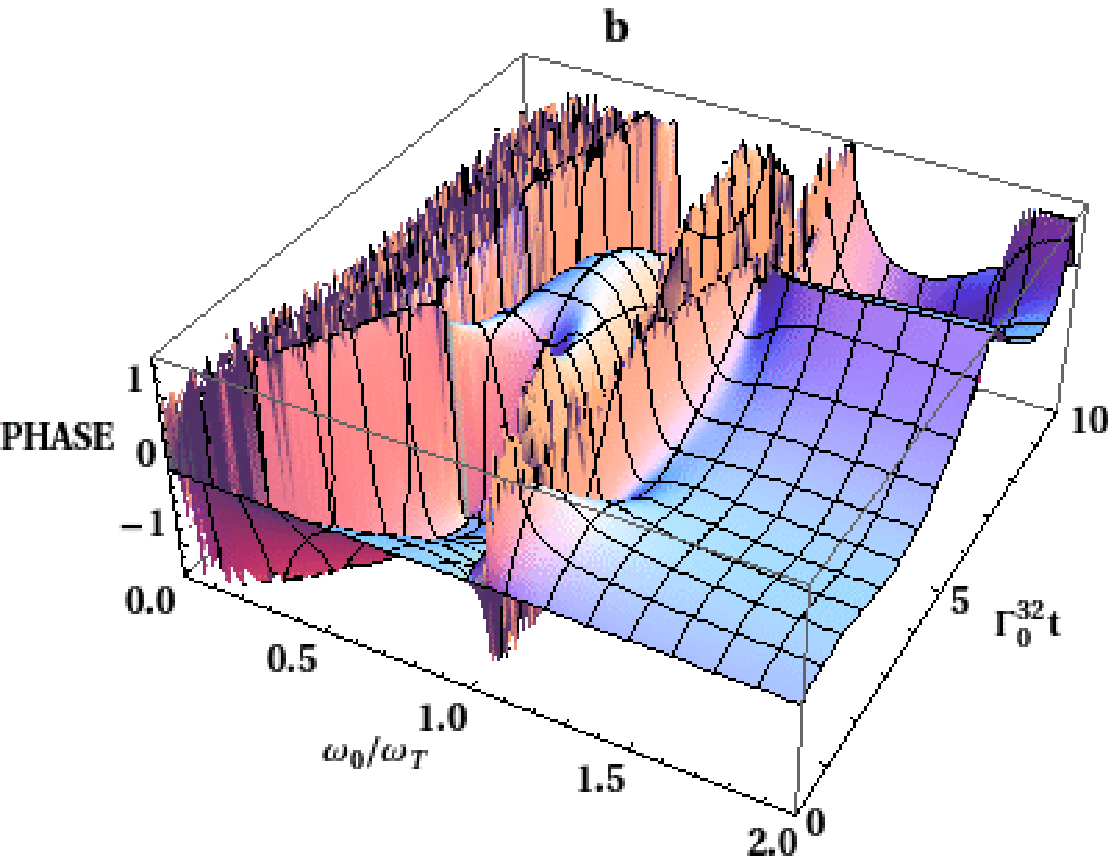}
\end{center}
\vspace{-0.5cm}
\caption{The same as Fig. 4 but when taking into account the effect of both $\Gamma^{32}$ as well as $\delta\omega_{0}$}
\label{Fig.10}
\end{figure}
\vspace{-0.3cm}
Figures 10-15 illustrate the dependence of the populations, topological Berry phase, $\phi_{t}$, and the fluctuations in $X$ and $Y$ on line shift $\delta\omega_{0}$ as well as the decay rate $\Gamma_{0}^{32}$ of an atom surrounded by a planar dielectric half-space with various values of the bandwidth parameter $\gamma$. The figures reveal that with the decrease of bandwidth parameter $\gamma$, specially outside the band gap region, where ; $\omega_{0}<\omega_{T}$, i.e., $\varepsilon(\omega_{0})>0$, the surface-guided waves, due to Drude-Lorentz model, give rise to strong enhancement of functions' oscillations to the extent that the lines strongly overlap in all  time stages regardless of the population decay. 
\begin{figure}[tpbh]
\noindent
\begin{center}
\includegraphics[width=0.32\linewidth]
{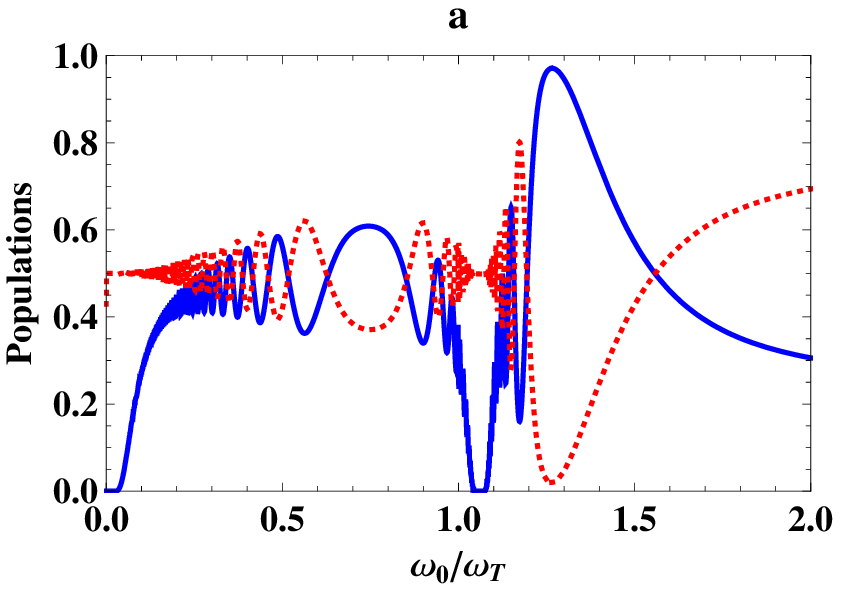}
\includegraphics[width=0.32\linewidth]
{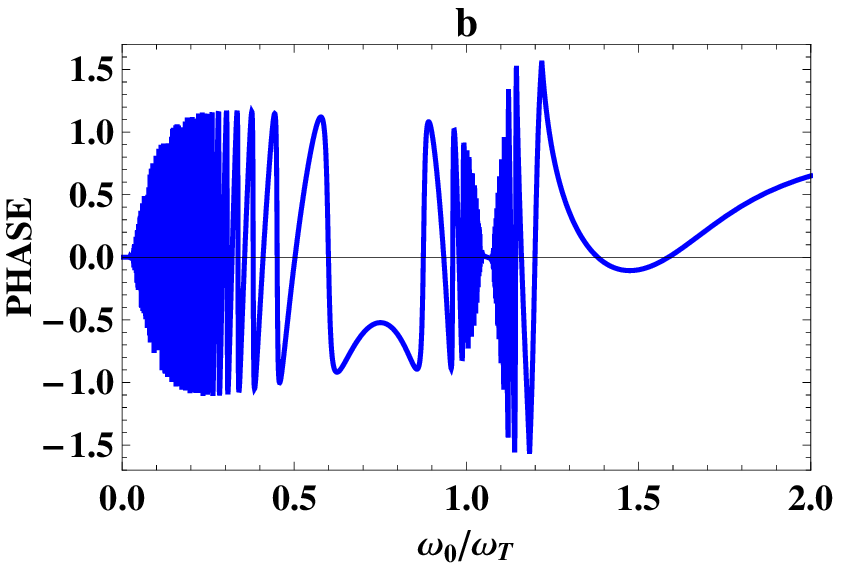}
\includegraphics[width=0.32\linewidth]
{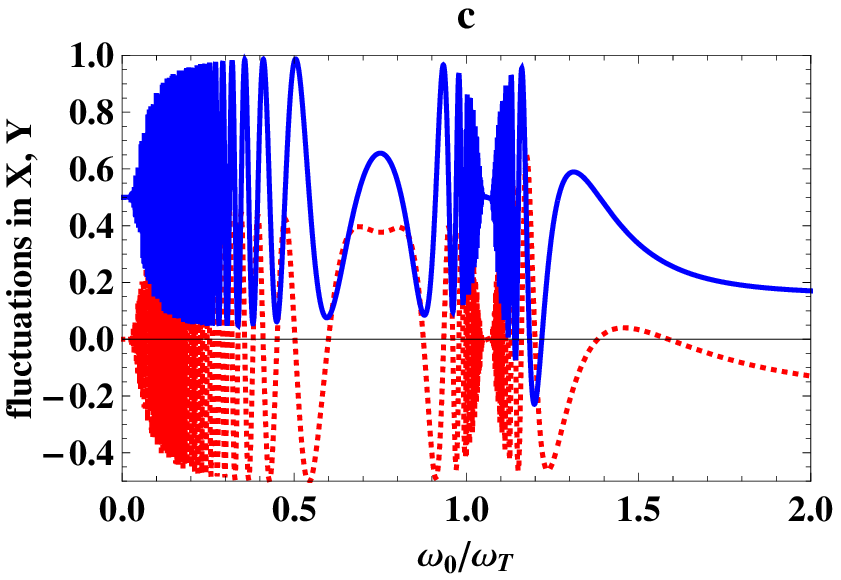}
\end{center}
\vspace{-0.7cm}
\caption{The same as Fig. 5 but when taking into account the effect of both $\Gamma^{32}$ as well as $\delta\omega_{0}$}
\label{Fig.11}
\end{figure}
The same effect can be easily seen in $X$ and $Y$ with particular emphasis on that $X$ reveals oscillations in all positive half plane outside the band gap region while goes once to the negative half plane inside the band gap region before accomplishes its decay. The function $Y$ oscillates around zero with approximately the same behavior of $X$. Also, for sufficiently small values of the bandwidth parameter,$\gamma$, below the band gap,  energy transfer between the atom and the medium occurs rapidly to the extent that the populations of the levels $\mid 2\rangle$ and $\mid 1\rangle$ overlap except for small course of time, while energy transfer strongly enhanced above the band gap region, which clarifies the high maxima and minima of $\phi_{t}$ [see Fig. 11]. On increasing $\gamma$, populations overlap breaks down gradually and, at end,  energy transfer stops with high values of $\gamma$, where the atom preserves its lower state and energy transfer occurs only above band gap region with small rate as can be seen by comparing Figs. 13 and 15.  This means a small separation angles can be produced. All these results support our idea that selection of the atomic state leads to a higher degree of nonclassical effects in the resulting radiation field. 
\begin{figure}[tpbh]
\noindent
\begin{center}
\includegraphics[width=0.32\linewidth]
{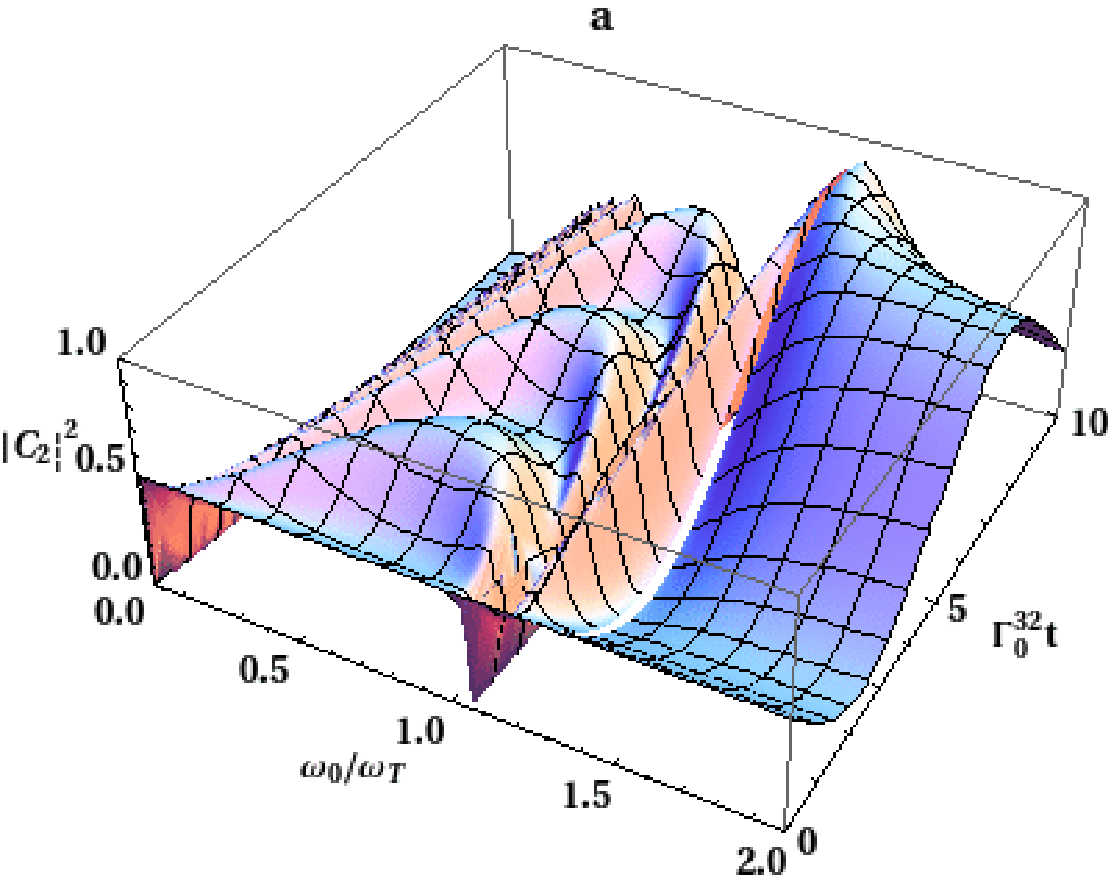}
\hspace{1.0cm}
\includegraphics[width=0.32\linewidth]
{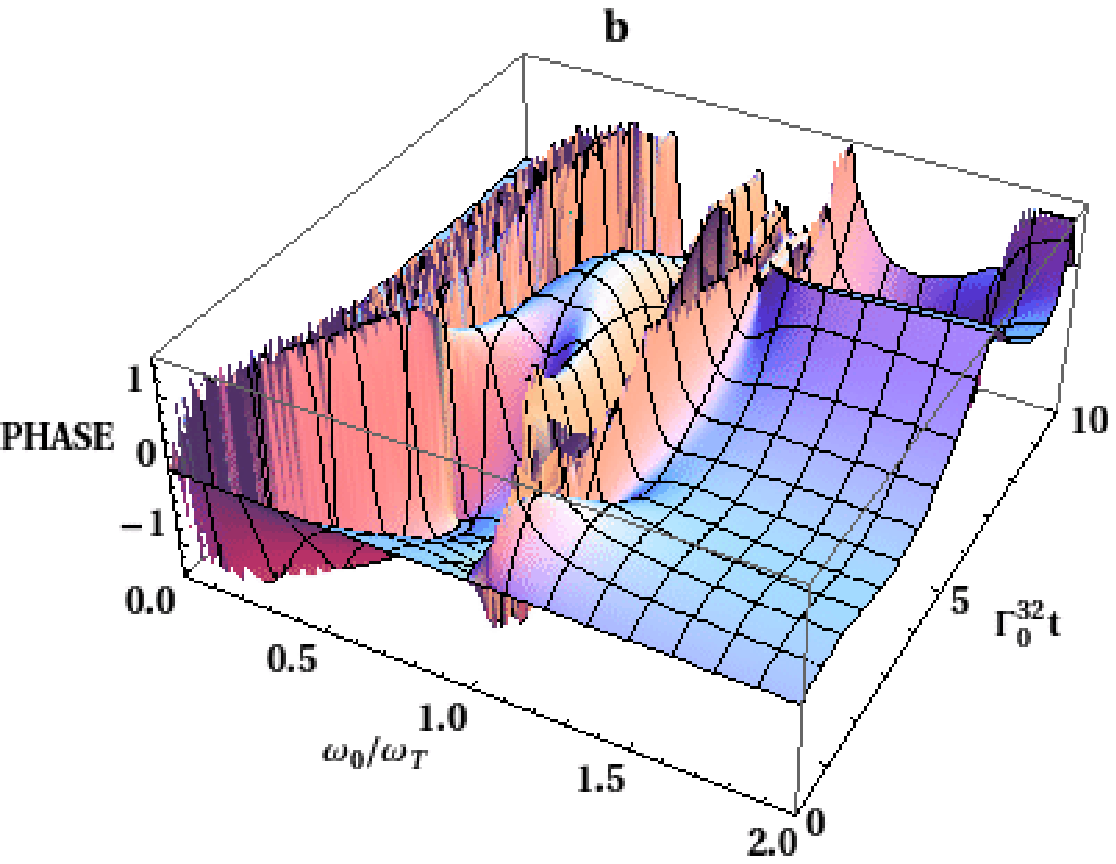}
\end{center}
\vspace{-0.5cm}
\caption{The same as Fig. 10 but for  $\gamma=10^{-2}\omega_{T}$}
\label{Fig.12}
\end{figure}
\begin{figure}[tpbh]
\noindent
\begin{center}
\includegraphics[width=0.32\linewidth]
{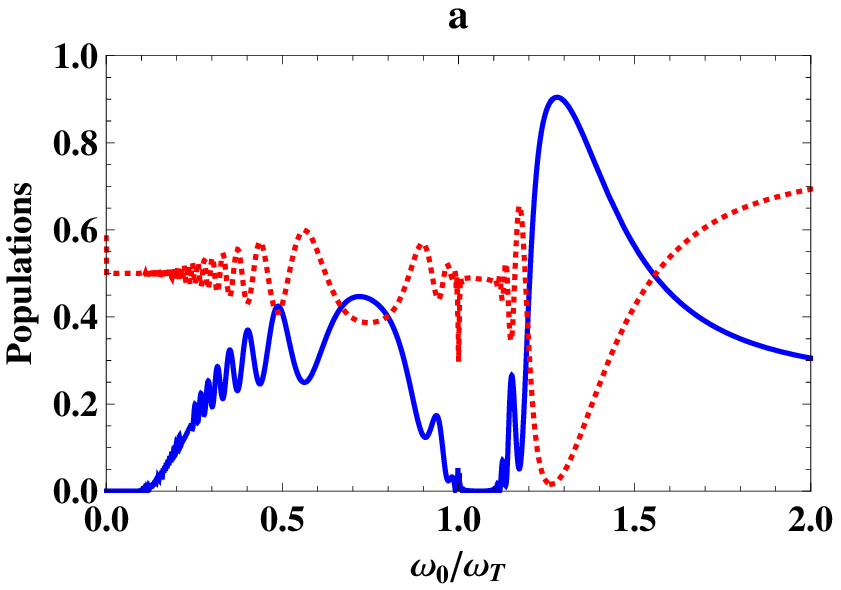}
\includegraphics[width=0.32\linewidth]
{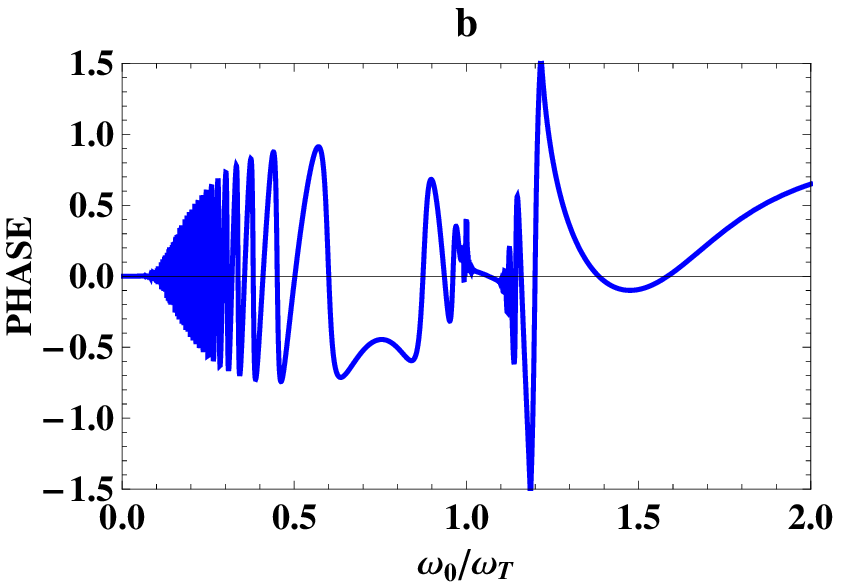}
\includegraphics[width=0.32\linewidth]
{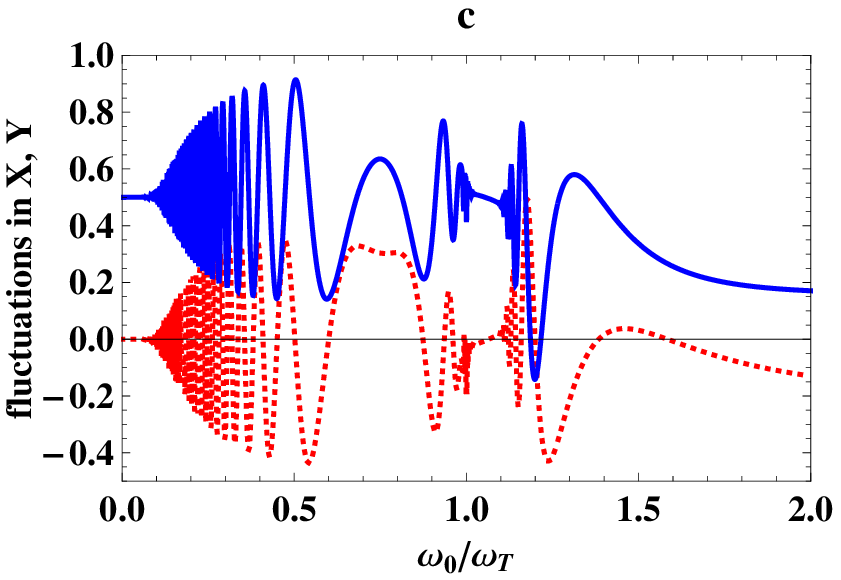}
\end{center}
\vspace{-0.7cm}
\caption{The same as Fig. 11 but for  $\gamma=10^{-2}\omega_{T}$}
\label{Fig.13}
\end{figure}
\begin{figure}[tpbh]
\noindent
\begin{center}
\includegraphics[width=0.32\linewidth]
{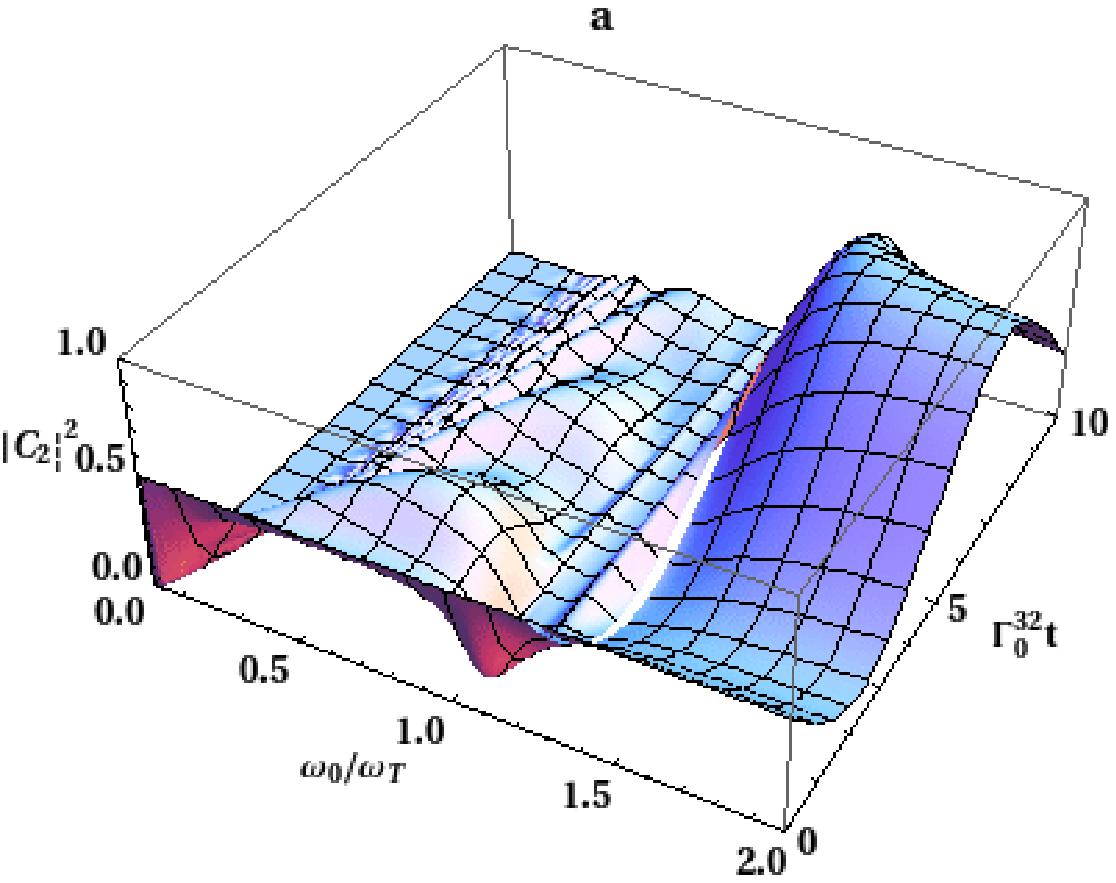}
\hspace{1.0cm}
\includegraphics[width=0.32\linewidth]
{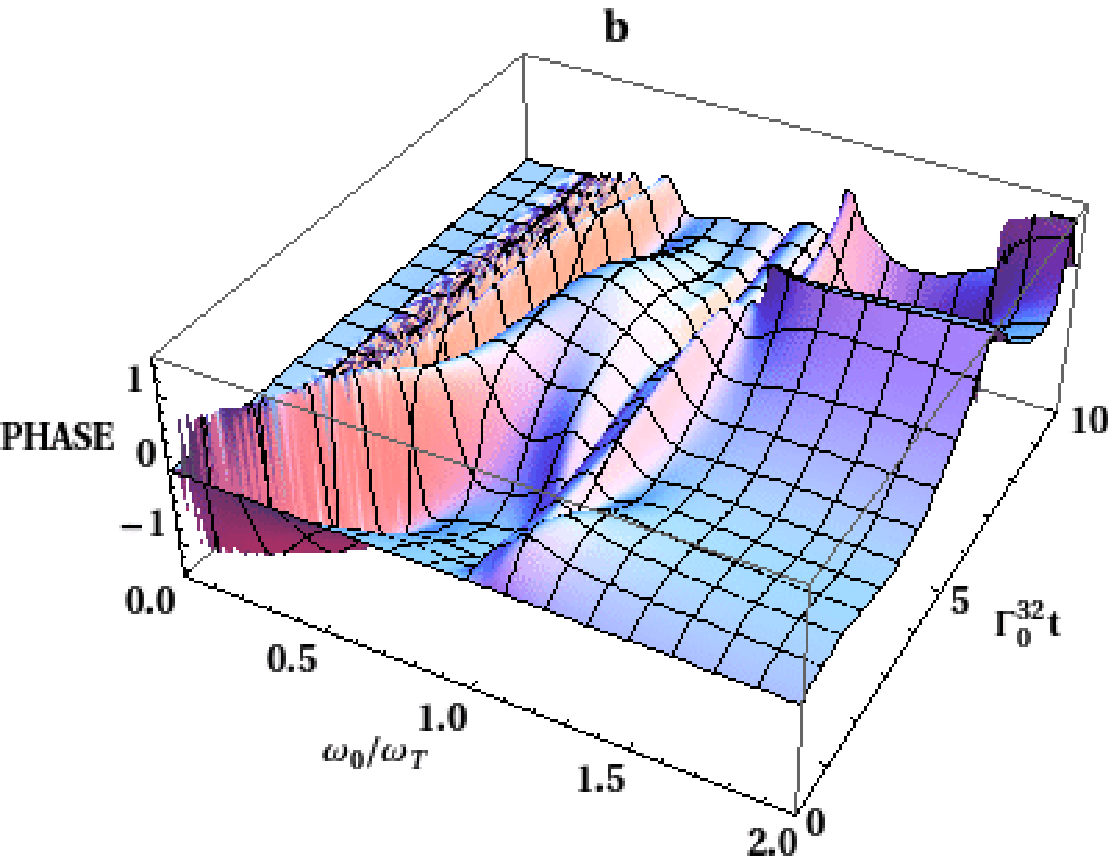}
\end{center}
\vspace{-0.5cm}
\caption{The same as Fig. 10 but for $\gamma=10^{-1}\omega_{T}$}
\label{Fig.14}
\end{figure}
\begin{figure}[tpbh]
\noindent
\begin{center}
\includegraphics[width=0.32\linewidth]
{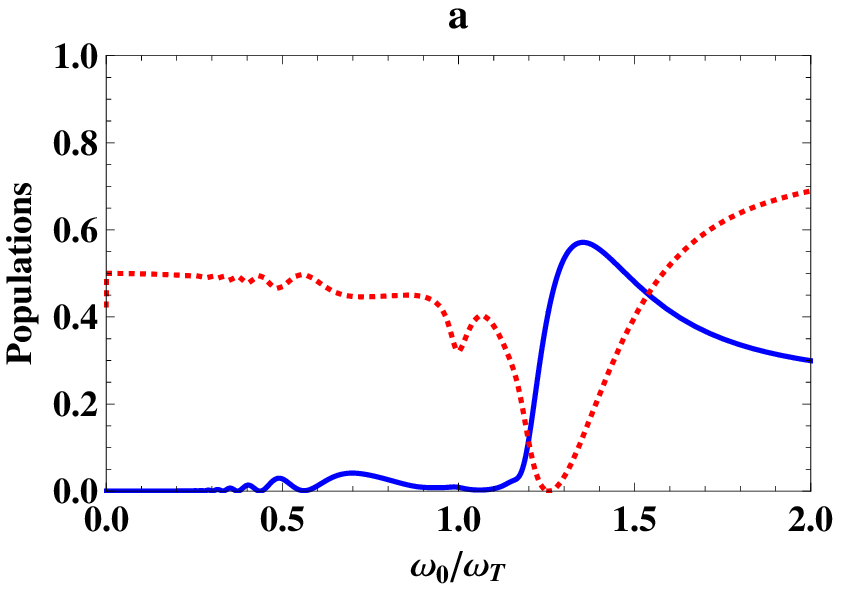}
\includegraphics[width=0.32\linewidth]
{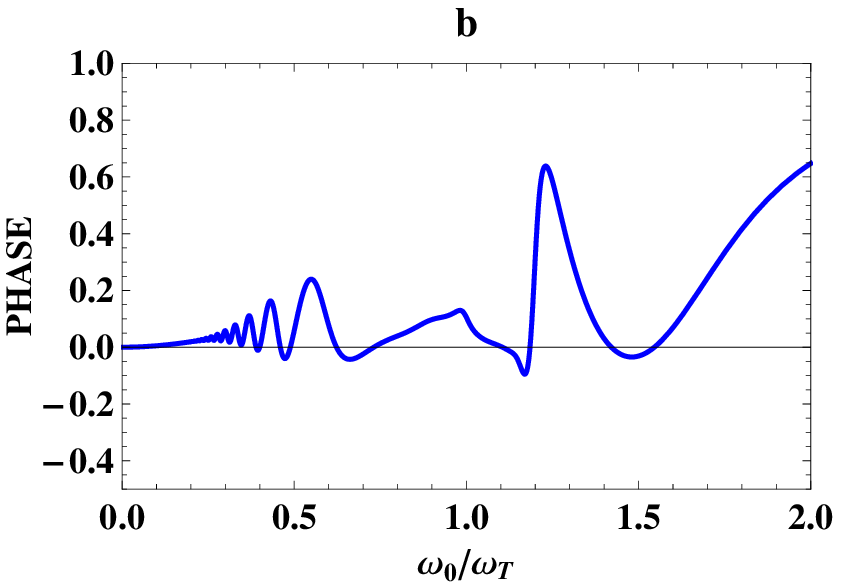}
\includegraphics[width=0.32\linewidth]
{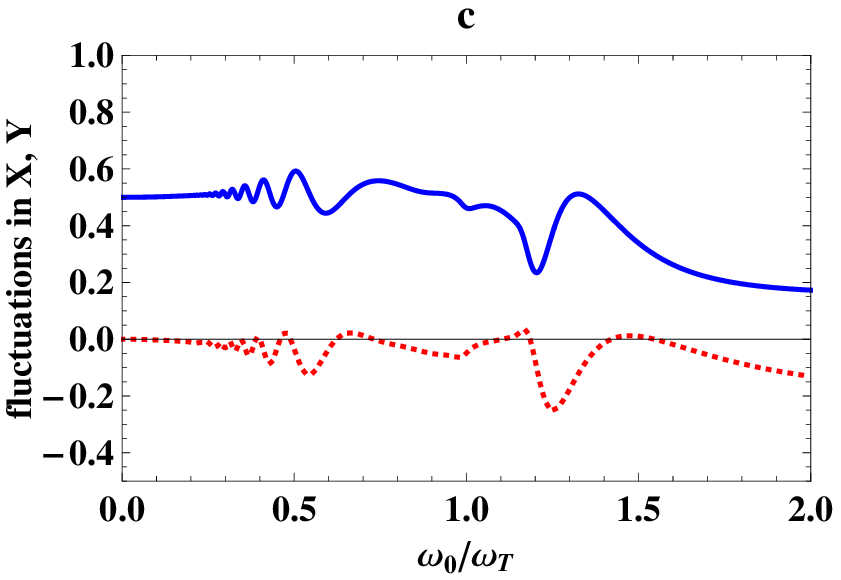}
\end{center}
\vspace{-0.7cm}
\caption{The same as Fig. 11 but for $\gamma=10^{-1}\omega_{T}$}
\label{Fig.15}
\end{figure}
%
%
\begin{figure}[tpbh]
\noindent
\begin{center}
\includegraphics[width=0.32\linewidth]
{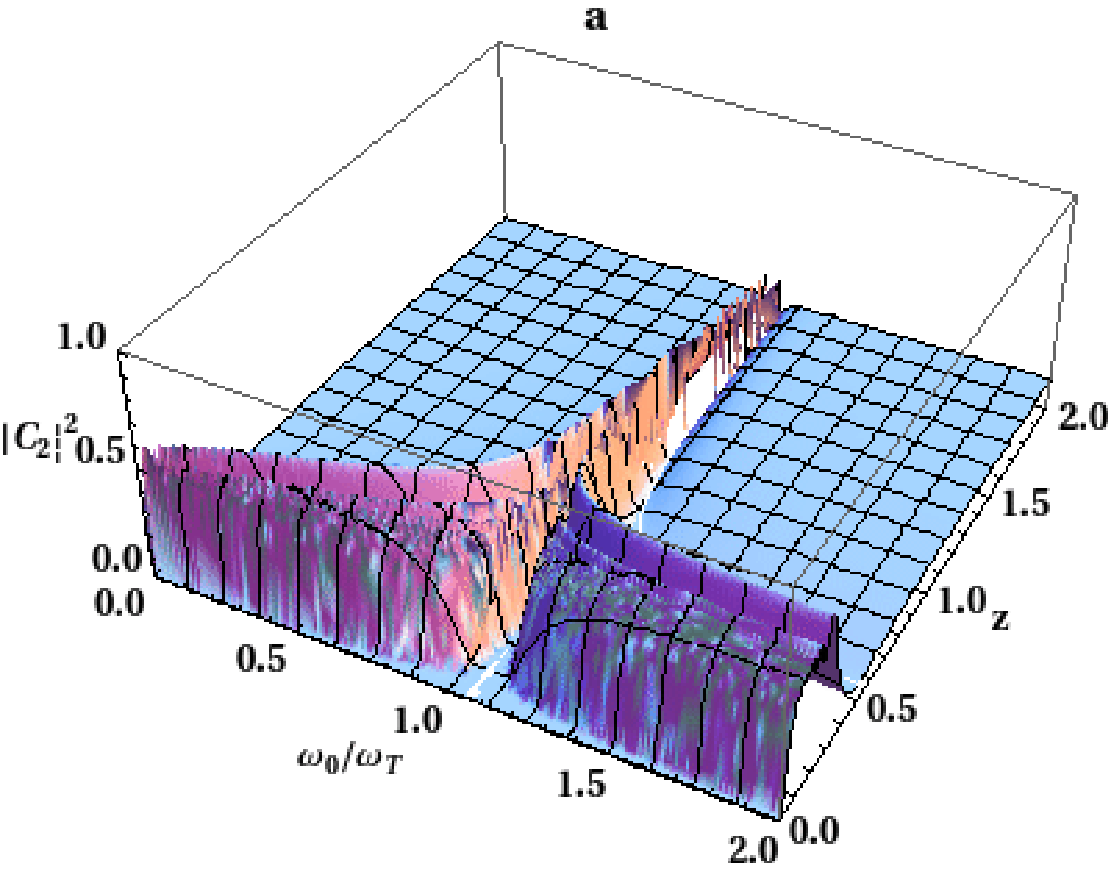}
\hspace{1.0cm}
\includegraphics[width=0.32\linewidth]
{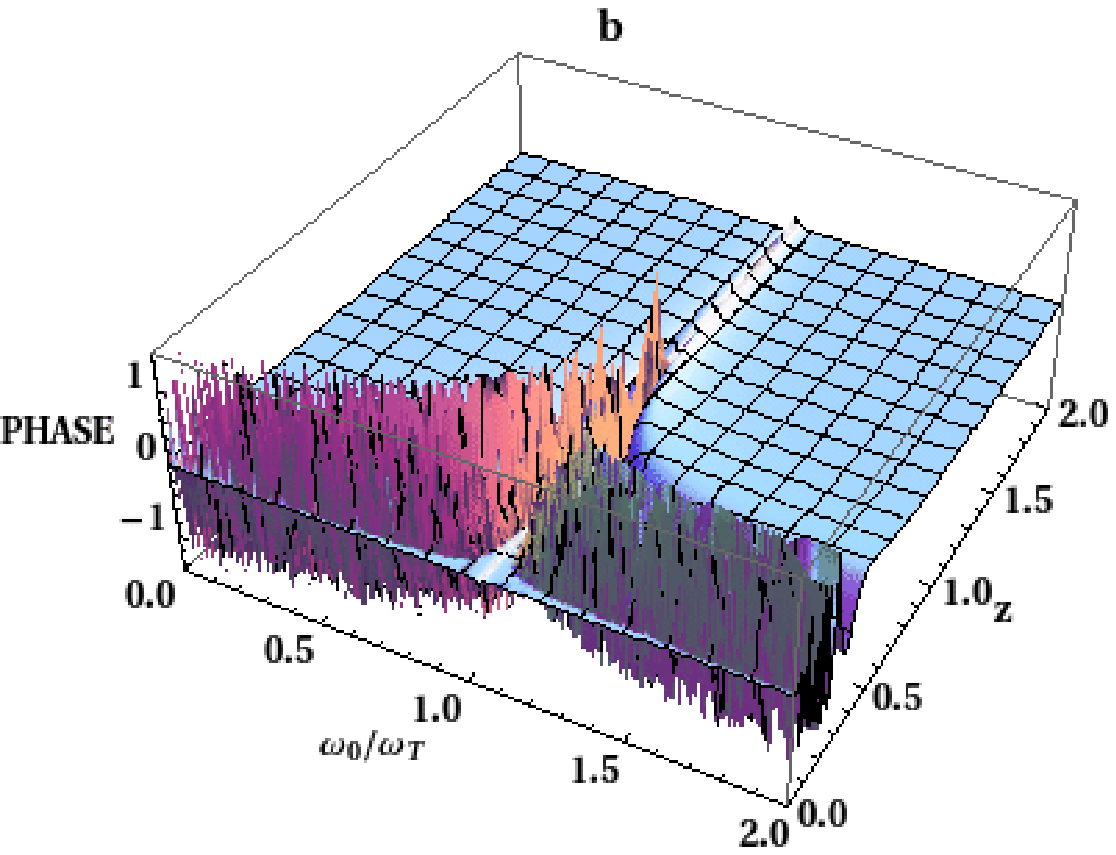}
\end{center}
\vspace{-0.5cm}
\caption{Mesh plot of $\mid 2\rangle$ state population (a) and the topological Berry phase $\phi_{t}$ (b) as functions of the transition frequency  $\omega_{0}/\omega_{T}$ and $z_{A}$ near a planar dielectric half-space for an $x$-oriented transition dipole moment, with
 $\omega_{p}=0.5\omega_{T}$ and $\gamma=10^{-3}\omega_{T}$}
\label{Fig.16}
\end{figure}
Now we turn our attention to the effect of the distance $z_A$ of the atom from the surface of the plate. Examples of the populations, topological Berry phase, $\phi_{t}$, and fluctuations in $X$, and $Y$, as functions of $z_A$ in the vicinity of a surface-guided field are shown in Figs. 16 - 21, where a single-resonance Drude-Lorentz-type dielectric has been assumed. One can see that the oscillations collapse after a few Rabi periods - for small interval below the band gap- and after an interval of time topological Berry phase $\phi_{t}$ vanishes, which means an in phase photon transition. In this case, an energy transfer between the atom and the surface occurs for distances short from the surface leaving the atom dexcited regardless of its distance from the surface due to the deep penetration of the emitted photon into the surface. This means, physically, that, in this case, the atomic transition frequency is relatively near medium resonance, so that the imaginary part of the permittivity becomes relatively big which enhances medium absorption. In addition, inside the band gap region, $\phi_{t}$ preserves fixed value - even when energy transfer occurs for sufficiently small values of $\gamma$ [Fig. 16] - regardless of the atom being far or close to the surface plate. As $\gamma$ increases Rabi periods shortened gradually where the general behavior preserved. So one can clearly see that the topological phase $\phi_{t}$ is affected strongly by the presence of the half space for atom-half-space separations that are greater than the medium oscillation wave length $\lambda_{T}$ ($\phi_{t}\approx 0$ for $z_A\geqslant 0.5\lambda_{T}$ as the figures suggest). These results means that by adjusting the bandwidth parameter $\gamma$, Rabi strength of the driving classical, $\Omega_{pmp}$ and the distance between the atom and the surface, $z_A$, quantum mechanically, we can with full success, through adiabatic quantum process, expect that the cyclic evolution of wavefunction yields the original state plus a zero phase shift, namely, cyclic evolution of wavefunction develops on the same track of the original state. In other words, the angular separation between the wavefunction evolution and the original state dies, that is of strong applications in construction of the universal quantum logic gates. \\
\begin{figure}[tpbh]
\noindent
\begin{center}
\includegraphics[width=0.32\linewidth]
{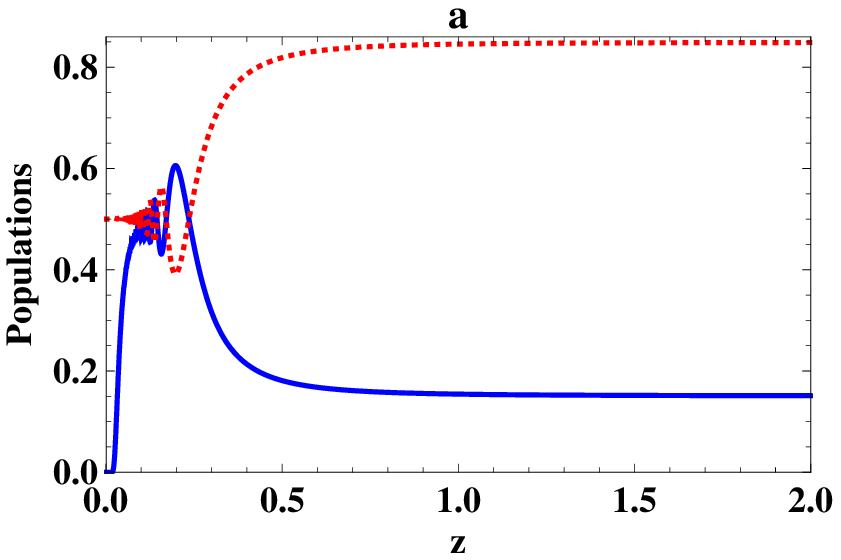}
\includegraphics[width=0.32\linewidth]
{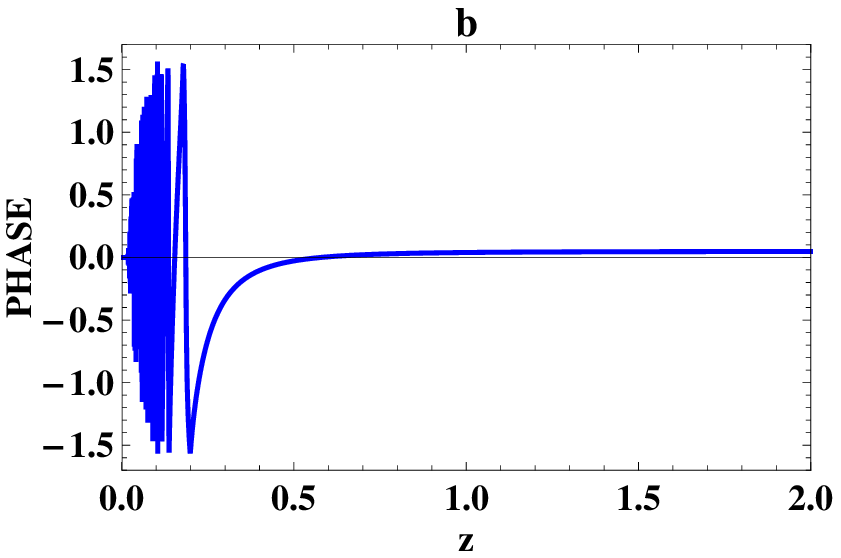}
\includegraphics[width=0.32\linewidth]
{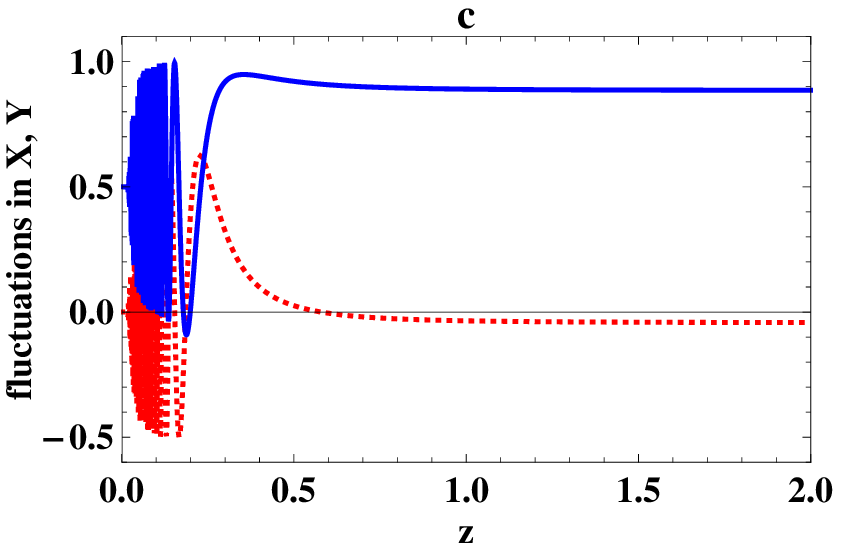}
\end{center}
\vspace{-0.7cm}
\caption{Cross section of Fig. 16 as well as fluctuations in $X$ and $Y$  (c), when $\Gamma_{0}^{32} t=2.0$, $\omega_{0}=0.5\omega_{T}$ and  $\gamma=10^{-3}\omega_{T}$,  }
\label{Fig.17}
\end{figure}
In fact, in quantum computation, operations are performed by means of single-qubit and multiple-qubit quantum logic gates \cite{BoEkZe01, LeBe03}. One could use the present model to generate a C-NOT gate, i.e., universal quantum logic gate based on three-level atom. For polarising beam splitter (PBS), for an emitted photons,  PBS redirects vertically polarized photons ( in state $\mid V\rangle$)  without affecting horizontally polarized photons (in state $\mid H\rangle$) such that
a  C-NOT gate can be obtained as
\begin{equation}
 \label{E64}
\alpha\mid H\rangle_{1}+\beta\mid V\rangle_{2}\mapsto \alpha\mid H\rangle_{1{'}}+\beta\mid V\rangle_{2{}'},
\end{equation}
where, for phase plate (PP) which adds an overall phase factor to the state of the incoming photon such that
\begin{equation}
 \label{E65}
\mid \psi\rangle_{1}\mapsto e^{i\phi}\mid \psi\rangle_{1{'}},
\end{equation}
with phase factor $\phi=0$, for present case. This can be accomplished my moving the atom away from the surface of the half-space. Note that there are values of the atom-surface separation at which the topological Berry phase is typically trivial [Figs. 17 and 18].
\begin{figure}[tpbh]
\noindent
\begin{center}
\includegraphics[width=0.32\linewidth]
{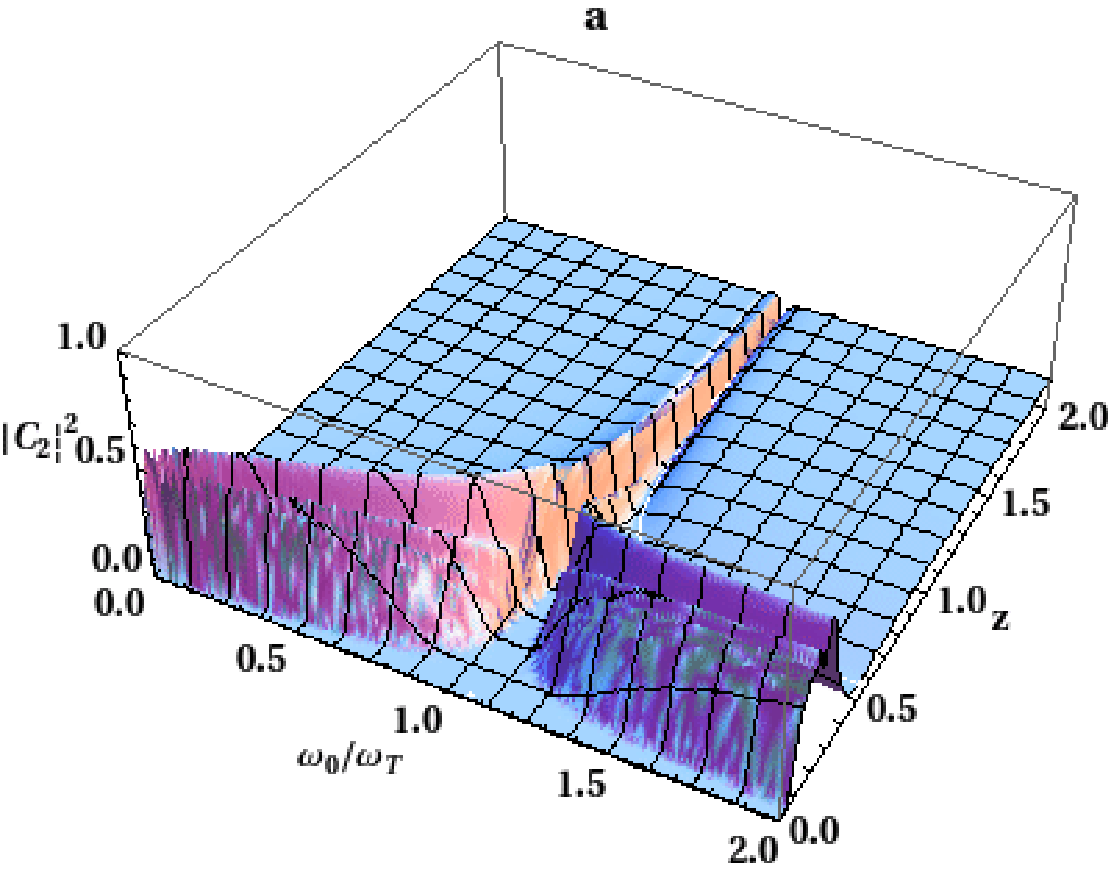}
\hspace{1.0cm}
\includegraphics[width=0.32\linewidth]
{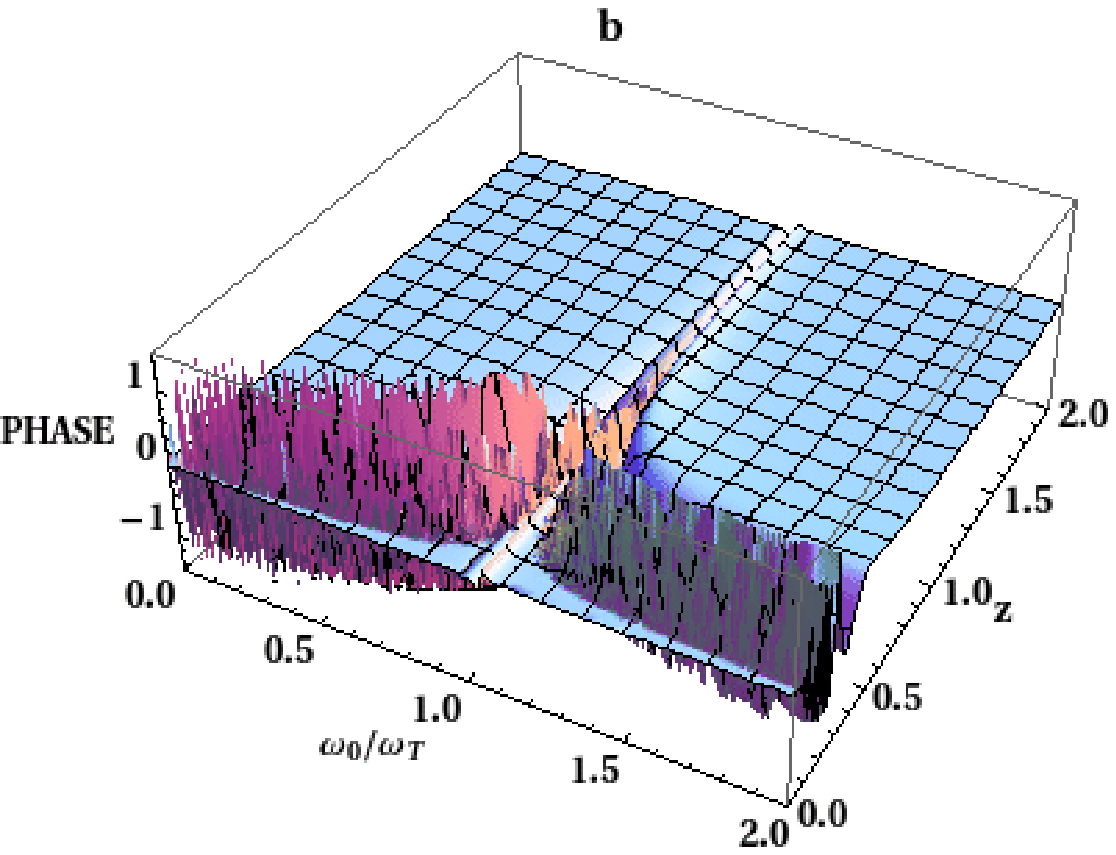}
\end{center}
\vspace{-0.5cm}
\caption{The same as Fig. 16 but for  $\gamma=10^{-2}\omega_{T}$.}
\label{Fig.18}
\end{figure}
\begin{figure}[tpbh]
\noindent
\begin{center}
\includegraphics[width=0.32\linewidth]
{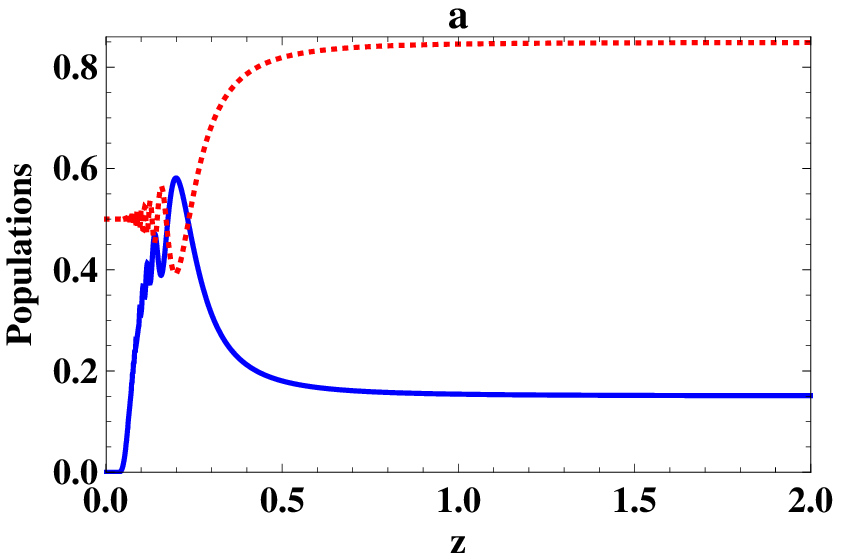}
\includegraphics[width=0.32\linewidth]
{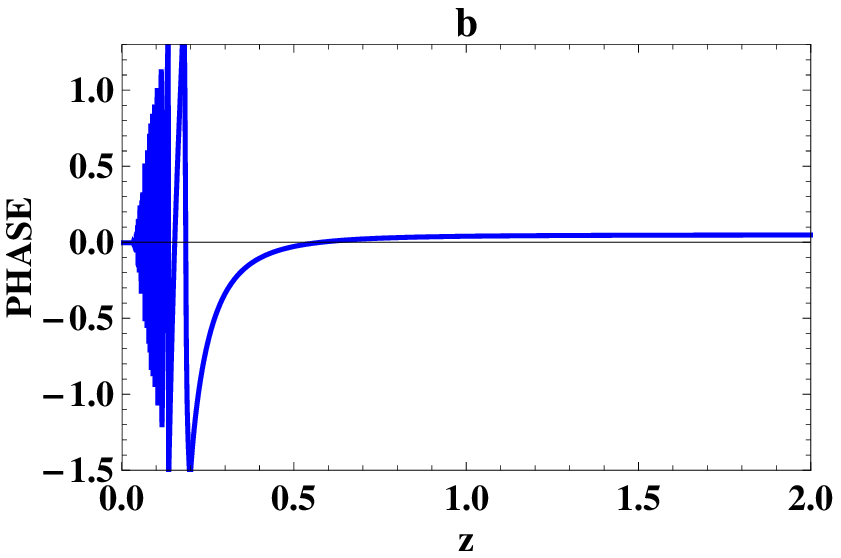}
\includegraphics[width=0.32\linewidth]
{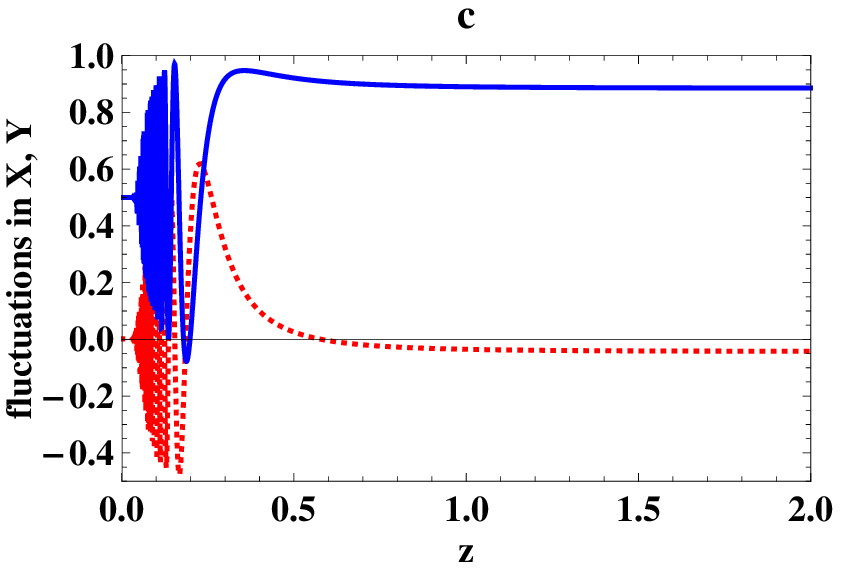}
\end{center}
\vspace{-0.7cm}
\caption{The same as Fig. 17 but for $\gamma=10^{-2}\omega_{T}$.}
\label{Fig.19}
\end{figure}
\begin{figure}[tpbh]
\noindent
\begin{center}
\includegraphics[width=0.32\linewidth]
{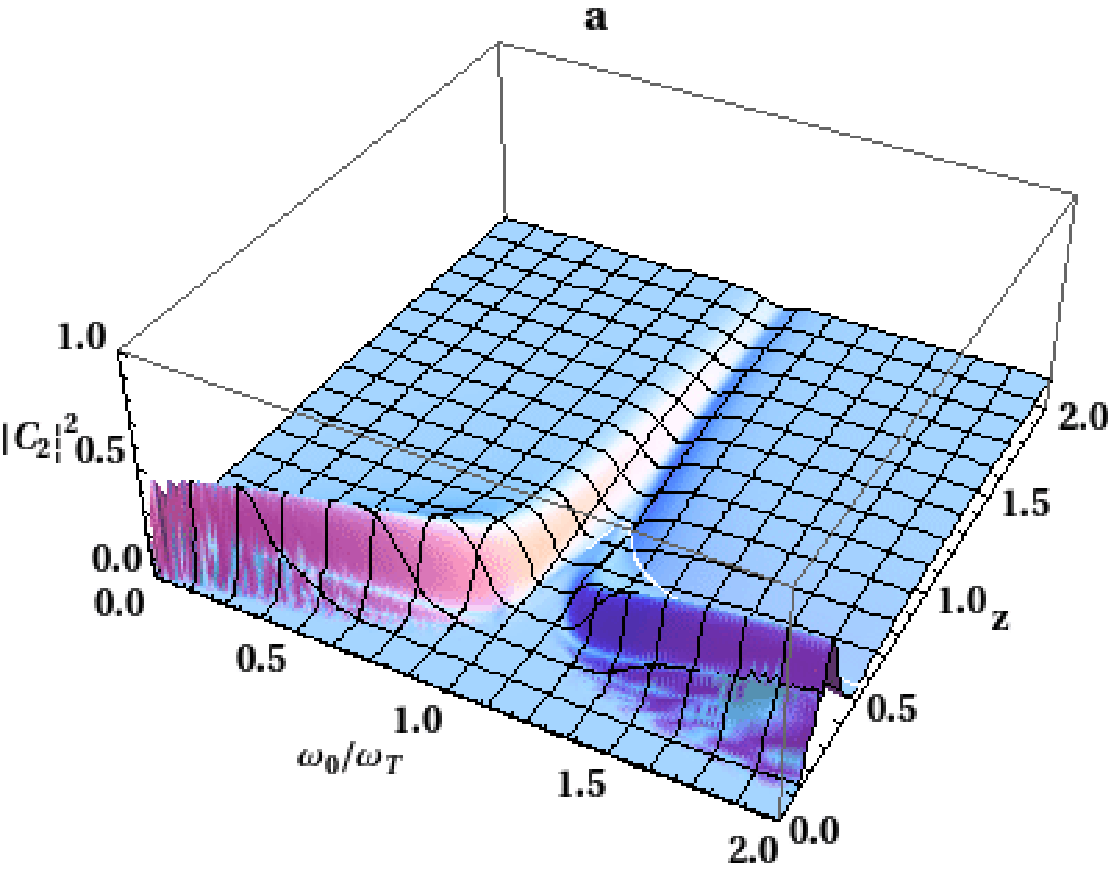}
\hspace{1.0cm}
\includegraphics[width=0.32\linewidth]
{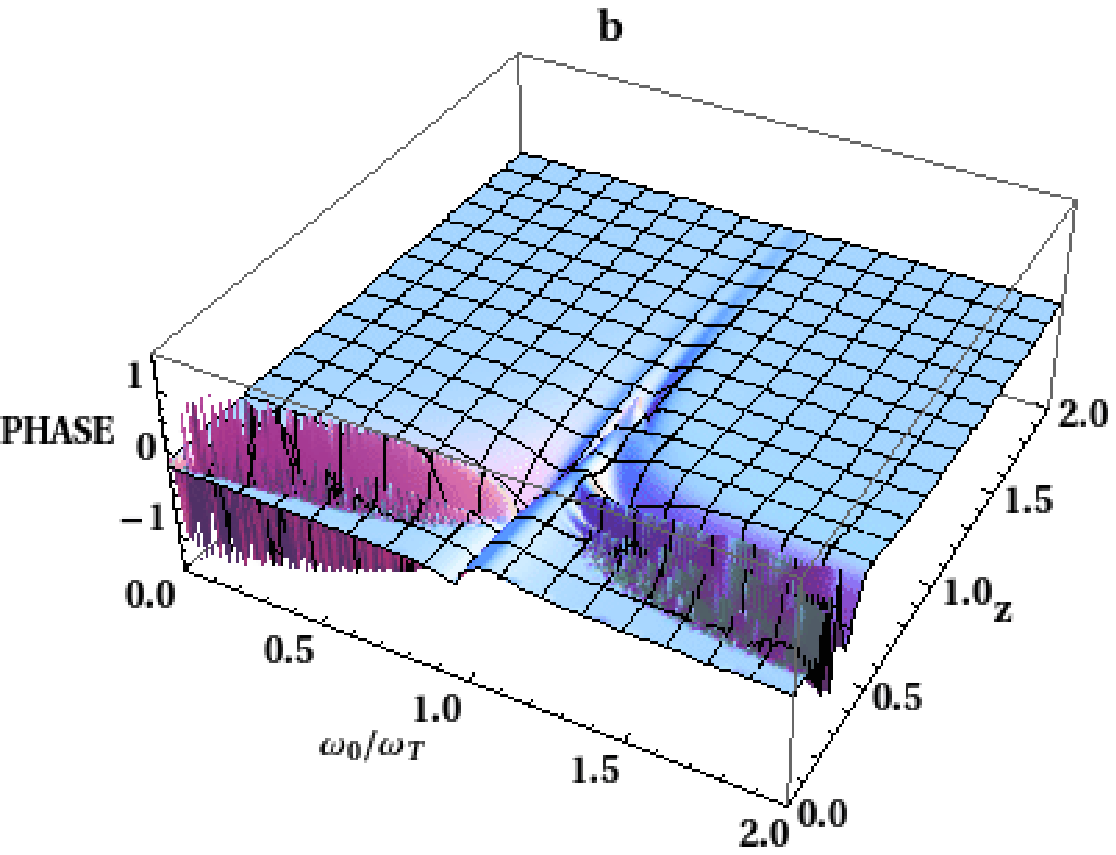}
\end{center}
\vspace{-0.5cm}
\caption{The same as Fig. 16 but for $\gamma=10^{-1}\omega_{T}$.}
\label{Fig.20}
\end{figure}
\begin{figure}[tpbh]
\noindent
\begin{center}
\includegraphics[width=0.32\linewidth]
{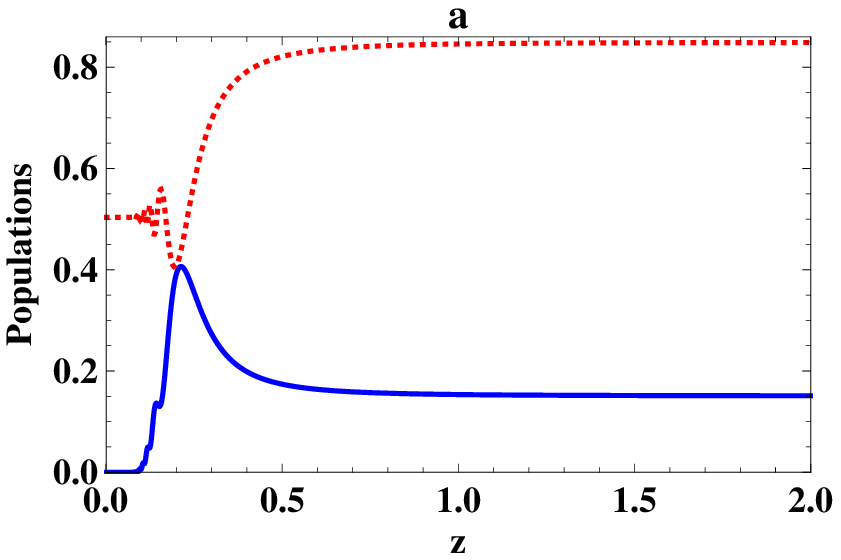}
\includegraphics[width=0.32\linewidth]
{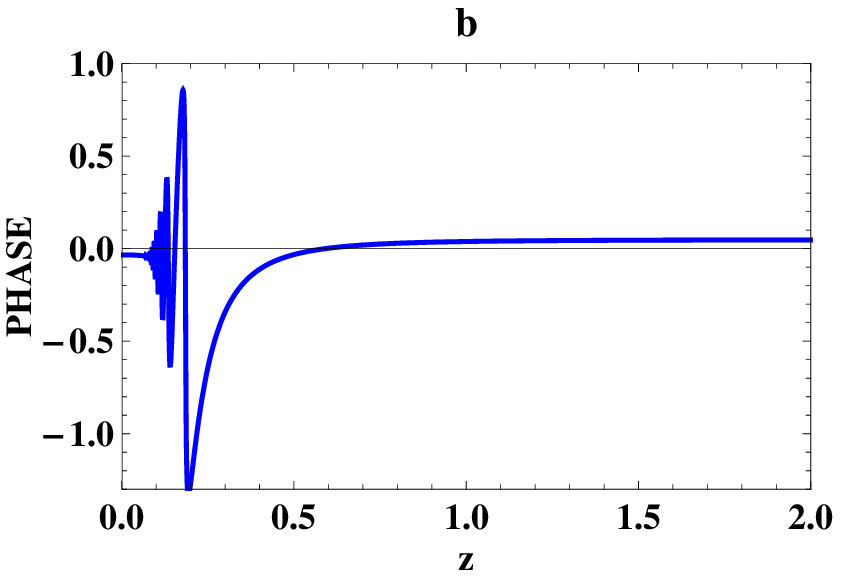}
\includegraphics[width=0.32\linewidth]
{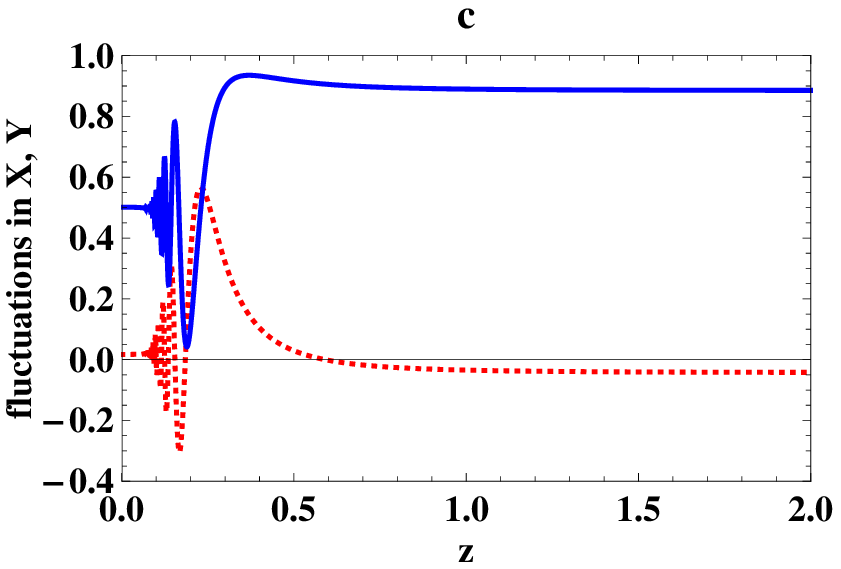}
\end{center}
\vspace{-0.7cm}
\caption{The same as Fig. 17 but for $\gamma=10^{-1}\omega_{T}$.}
\label{Fig.21}
\end{figure}
\section{SUMMARY AND CONCLUSIONS}
Within the frame of macroscopic QED in dispersing and absorbing media, we have investigated the topological Berry phase properties of the photon transition between two classically pumped levels of three-level atom interacting with the electromagnetic vacuum field in the presence of dispersing and absorbing material surroundings described by a spatially varying permittivity that is a complex function of frequency.  Using a source-quantity representation of the electromagnetic field, and starting from the exact minimal-coupling Hamiltonian (in the electric dipole approximation), we have performed the calculations for a one-dimensional Hamiltonian of the total system that is accordingly used in the calculation of the wavefunction, in a general form, that governs the system utilizing Schr\"{o}dinger equation. To gain further insight into the difference in the nature of the topological Berry phase function,
we presented a more subtle comparison to include the atom in free space. Without {\it ab initio} calculations, utilizing the physical properties of vacuum field evolution applied to the kernel function $K(t,t{}')$, we have derived the complex amplitudes of the classically pumped levels in free space, hence, the argument overlap between photon transition is also calculated applying the Weisskopf-Wigner approximation. On considering the atom near a dielectric body, we restricted our selves to a planar dielectric half-space with model permittivity of Lorentz type. Effects of various physical settings regarding the transition frequency, bandwidth parameter and the position of the atom, $z_A$, relative to the dielectric surface have been investigated. The results show that, in contrast to free space case in which the probability amplitudes of the classically pumped levels decay exponentially, the emitted light observed, and the topological Berry phase between photon transition can be controlled strongly by the presence of macroscopic bodies as follows: 
\begin{itemize}
 \item For relatively fixed small distance $z_A$, between the atom from and dielectric surface, specially, in the case that the line shift is canceled, the topological Berry phase exhibits maximum angles separation between the wavefunction evolution and original state in two positions, namely, before interaction is switched on and in the band gap region where the decay rate reaches its maximum. This behavior dominates accompanied with band-gap-envelope phase gets more broaden as bandwidth grows. When line shift is taken into consideration, the phase shows Rabi oscillations, specially below the band gap region, while these oscillations invoked below and above band gap with  bandwidth increases.
\item Moving the atom a way from the dielectric surface leads to phase angle evolution in straight line coincide with $x$-axis, which means an in-phase photon transition frequency. 
\item For small separation distance for atom placed in parallel alignment with respect to a half space, the dielectric surface-induced modification of the topological Berry phase is very different from our findings of atom placed in free space: A dielectric surface  reduces the angle separation between the wavefunction evolution and the original state, while free space leads to an enhancement. In other words, if the atom is placed sufficiently near the dielectric surface, its spontaneous decay may be suppressed, with the emission line being accordingly narrowed, which encourage energy transfer between the atom and the dielectric medium.
\item Further, in contrast to a classically pumped atom in free space, the exponential decay dynamics gives place to the non-exponential one if the atom moves away from the dielectric surface due to the coexistence effects of the dielectric body and classical pumping laser field. Thus, the atom-field coupling strength and the character of the spontaneous decay dynamics, respectively, may be controlled by changing the distance between the atom and the dielectric surface by means of a proper preparation of classical pumping laser field. 
\item At end, it should be stressed that transition frequency shifts experience significantly enhanced changes than in free space case.
\item Our study opens routes for new challenging applications of atomically systems as various sources of coherent light emitted by pumped atoms in dielectric surroundings. 
\item The (numerical) calculations performed for atoms near a half-space can of course be extended to other geometries of the material surroundings such as microspere or cylindrical objects.
\end{itemize}
\appendix
\section{THE HAMILTONIAN IN THE DIPOLE AND ROTATING WAVE APPROXIMATIONS}
\label{AppA}
Applying the minimal-coupling scheme (\ref{E1}), we may decompose the Hamiltonian (\ref{E1}) of the coupled system consisting of the atom and medium-assisted electromagnetic field as
\begin{equation}
\label{A1}
\hat{H}= \hat{H}_{F}+\hat{H}_{A}+\hat{H}_{AF}
\end{equation}
where
\begin{equation}
\label{A2}
\hat{H}_{F}=\int d^{3}{\bf r}\int_{0}^{\infty} d\omega~\hbar\omega~ \hat{\bf f}^{\prime\dag}({\bf r},\omega)~\hat{\bf f}^{\prime}({\bf r},\omega) 
\end{equation}
is the Hamiltonian of the electromagnetic field and the dielectric matter, 
\begin{equation}
\label{A3}
\hat{H}_{A}=\sum_{j}\frac{\hat{{\bf p}}_{j}^2}{2m_{j}}+\frac{1}{2}\int d^{3} {\bf r}\hat{\rho}_A ({\bf r})\hat{\varphi}_{A}({\bf r}),
\end{equation}
is the Hamiltonian of the atom, and
\begin{equation}
\label{A4}
\hat{H}_{AF}=-\sum_{j}\frac{Q_{j}}{m_{j}}\hat{{\bf p}}_{j}\cdot \hat{{\bf A}}(\hat{{\bf r}}_{j})+\int d^{3} {\bf r} \hat{\rho}_A ({\bf r}){ \hat{\varphi}}({\bf r}),
\end{equation}
is the interaction energy. Here we omitted the $\hat{{\bf A}}^2$ term as a result of the (quasi-)resonant atom-field interaction. In the electrc-dipole approximation the first term on the right-hand side of (\ref{A4}) simplifies to
\begin{equation}
\label{A5}
-\sum_{j}\frac{Q_{j}}{m_{j}}\hat{{\bf p}}_{j}\cdot \hat{{\bf A}}(\hat{{\bf r}}_{j})=-\sum_{j}\frac{ Q_{j}}{i\hbar}[\hat{{\bf r}}_{j},\hat{H}_{A}]\cdot \hat{{\bf A}}({\bf r}_{A})=-\frac{ 1}{i\hbar}[\hat{{\bf d}}_{A},\hat{H}_{A}]\cdot \hat{{\bf A}}({\bf r}_{A}),
\end{equation}
where 
\begin{equation}
\label{A6}
 \hat{{\bf d}}_{A}=\sum_{j} Q_{j}(\hat{{\bf r}}_{j}-{\bf r}_{A})=\sum_{j} Q_{j} \hat{{\bf r}}_{j},
\end{equation}
is the atomic dipole operator for a neutral atom with the nucleus being positioned at ${\bf r}_{A}$. Note, in (\ref{A5}) the dipole approximation has been employed by replacing $\hat{{\bf A}}({\bf r}_{j})\rightarrow\hat{{\bf A}}({\bf r}_{A})$ and the commutation relation  $\hat{{\bf p}}_{j}=-\frac{i m_{j}}{\hbar}[\hat{{\bf r}}_{j},\hat{H}_{A}]$, bearing in mind that in the Coulomb gauge $[\hat{{\bf p}}_{j},\hat{{\bf A}}]=0$, has been used.
\\
In the next step, the atomic Hamiltonian (for an isolated atom) satisfies the eigenvalue equation
\begin{equation}
\label{A7}
\hat{H}_{A}\mid k\rangle_{A}=\hbar\omega_k \mid k\rangle_{A},
\end{equation}
where $\hbar\omega_k$ are the energy eigenvalues and $\mid k\rangle$ are the eigenvectors. The states $\mid k\rangle$ span the Hilbert space of the atomic system. Since $\hat{H}_{A}$ is an observable, its basis vectors $\{\mid k\rangle\}$ form a complete orthonormal set as expressed by the closure (completeness) and orthonormality relations
\begin{equation}
\label{A8}
\sum_{k}\mid k\rangle_{A A} \langle k\mid =\hat{I},~~~~~~~~~~~~ _{A}\langle k\mid k{'} \rangle _{A}=\delta_{k k{}'},
\end{equation}
where $\hat{I}$ represents the identity operator, and $\delta_{kk{}'}$ is the Kronecker delta function ($\delta_{kk{}'}=1$ for $k = k{}'$ , and $\delta_{kk{}'}=0$ for $k\neq k{}'$ ). We define the atomic operator $\hat{S}_{A k k{}'}$ by
\begin{equation}
\label{A9}
 \hat{S}_{Akk{}'}\equiv \mid k\rangle_{A A} \langle k{}'\mid.
\end{equation}
Using Eq. (\ref{A8}), $\hat{H}_{A}$ can be written as
\begin{equation}
\label{A10}
\hat{H}_{A}=\sum_{k} \hbar\omega_k \hat{S}_{Akk}.
\end{equation}
Similarly, on using Eq. (\ref{A8}) twice, we can represent the atomic dipole-moment $\hat{{\bf d}}_{A}$ as
\begin{equation}
\label{A11}
 \hat{{\bf d}}_{A}=\sum_{k,k{}'}\hat{{\bf d}}_{A kk{}'} \hat{S}_{Akk{}'} ,
\end{equation}
with
\begin{equation}
\label{A12}
 \hat{{\bf d}}_{A kk{}'}=_{A}\langle k\mid \hat{{\bf d}}_{A} \mid k{}'\rangle_{A}.
\end{equation}
If we apply the same procedure on Eq. (\ref{A5}), we arrive at
\begin{equation}
\label{A13}
-\frac{ 1}{i\hbar}[\hat{{\bf d}}_{A},\hat{H}_{A}]\cdot \hat{{\bf A}}({\bf r}_{A})=i\sum_{k,k{}'} \omega_{k{}'k}\hat{S}_{Akk{}'}~\hat{{\bf d}}_{A k{}'k}\cdot \hat{{\bf A}}({\bf r}_{A});~~~~~~~~~~~~ \omega_{kk{}'}=\omega_{k}-\omega_{k{'}}=-\omega_{k{}'k},
\end{equation}
This equation applies to a general $k$-level atom interacting with a quantized electromagnetic field in the electric dipole approximation. In the following we will restrict our attention to a three-level system. A three-level atom (consisting of the ground level $\mid 3\rangle$ and excited levels $\mid 1\rangle$ and $\mid 2\rangle$) can be in one of three distinct ( cascade; $\Xi$, lambda; $\Lambda$ and $V$) configurations. Thus, for each of the three configurations there is a direct dipole transitions is forbidden and two dipole-allowed transitions. Restricting our attetion to $\Lambda$-type, direct dipole transition is forbidden between levels $\mid 1\rangle$ and $\mid 3\rangle$. With the property of $\hat{{\bf d}}_{A kk{}'}$ where, for states of the same parity $\hat{{\bf d}}_{A kk}=0$, bearing in mind that $\hat{{\bf d}}_{A kk{}'}$ is Hermitian; $\hat{{\bf d}}_{A kk{}'}=\hat{{\bf d}}_{A k{}'k}^{\ast}$,  equation (\ref{A5}), with use of Eq. (\ref{A13}), becomes
\begin{equation}
\label{A14}
-\sum_{j}\frac{Q_{j}}{m_{j}}\hat{{\bf p}}_{j}\cdot \hat{{\bf A}}(\hat{{\bf r}}_{j})
 =-i\omega_{23}\biggl[ \hat{S}_{A23}~\hat{{\bf d}}_{A 23}-\hat{S}_{A32}~\hat{{\bf d}}_{A 32}\biggl]\cdot \hat{{\bf A}}({\bf r}_{A})
\end{equation}
In the Coulomb gauge, the vector and scaler potentials are given, respectively, by
\begin{equation}
\label{A15}
 \hat{{\bf A}}({\bf r})=\int_{0}^{\infty} d\omega \frac{1}{i\omega}\hat{{\bf E}}^{\perp}({\bf r},\omega)+ \mathrm{H.c}
\end{equation}
\begin{equation}
\label{A16}
 -\nabla\hat{\varphi}({\bf r})=\hat{{\bf E}}^{\parallel}({\bf r})=\int_{0}^{\infty} d\omega \frac{1}{i\omega}\hat{{\bf E}}^{\parallel}({\bf r},\omega)+ \mathrm{H.c}
\end{equation}
where the transverse (longitudinal) electric field operator
\begin{equation}
\label{A17}
 \hat{{\bf E}}^{\perp(\parallel)}({\bf r})=\int d^{3} {\bf r}{}'\delta^{\perp(\parallel)}({\bf r}-{\bf r}{}')\cdot \hat{{\bf E}}({\bf r}{}'),
\end{equation}
with $\delta^{\perp}({\bf r})$ and $\delta^{\parallel}({\bf r})$ being the transverse and longitudinal $\delta$-functions respectively.
Recalling Eqs. (\ref{A15} -\ref{A17}), with the application of the rotating wave approximation, and after short calculations Eq. (\ref{A5}) reads
\begin{equation}
\label{A18}
 -\sum_{j}\frac{Q_{j}}{m_{j}}\hat{{\bf p}}_{j}\cdot \hat{{\bf A}}(\hat{{\bf r}}_{j})=-\hat{{\bf d}}_{A}\cdot\hat{{\bf E}}^{\parallel}({\bf r}_{A})
  \simeq-\Big(\hat{S}_{A23} \hat{{\bf E}}^{\parallel(+)}({\bf r}_{A})\cdot ~\hat{{\bf d}}_{A 23}+\mathrm{H.c.}\Big)
\end{equation}
where the total electric field is given by Eq. (\ref{E4}) with the transverse (longitudinal) electric field operator reads
\begin{equation}
\label{A19}
 \hat{{\bf E}}^{\perp(\parallel)}({\bf r}_{A})=\int_{0}^{\infty} d\omega \hat{{\bf E}}^{\perp(\parallel)}({\bf r}_{A},\omega)
\end{equation}
 and $\omega=\omega_{23}$ is set in the integral.\\
In order to deal with the second term in $\hat{H}_{AF}$, we expand $\hat{\rho}_{A}({\bf r})$ in a multi-polar form and retain only the first non-vanishing term
\begin{equation}
 \hat{\rho}_{A}({\bf r})=\sum_{j} Q_{j}\delta({\bf r}-{\bf r}_{A})-\nabla\cdot \Big(\delta({\bf r}-{\bf r}_{A})\sum_{j} Q_{j}({\bf r}_{j}-{\bf r}_{A})\Big)\nonumber
\end{equation}
\begin{equation}
=-\nabla\cdot \delta({\bf r}-{\bf r}_{A})\hat{{\bf d}}_{A}\simeq -\int d^{3} {\bf r} \Big[\nabla\cdot \delta({\bf r}-{\bf r}_{A})\hat{{\bf d}}_{A}\Big]{ \hat{\varphi}}({\bf r})\nonumber
\end{equation}
\begin{equation}
\label{A20}
= \int d^{3} {\bf r} [ \delta({\bf r}-{\bf r}_{A})\hat{{\bf d}}_{A}]\cdot\nabla{ \hat{\varphi}}({\bf r})=\hat{{\bf d}}_{A}\cdot\nabla{ \hat{\varphi}}({\bf r}_{A})=-\hat{{\bf d}}_{A}\cdot\hat{{\bf E}}^{\parallel}({\bf r}_{A})
\end{equation}
where integration by parts and Eq. (\ref{A16}) have been employed for deriving the second and the third equation, respectively. 
Recalling Eq. (\ref{A11}), restricting our attention to $\Lambda$-type with the same properties applied to (\ref{A14}), we arrive at
\begin{equation}
\label{A21}
\int d^{3} {\bf r} \hat{\rho}_A ({\bf r}){ \hat{\varphi}}({\bf r})=-\hat{{\bf d}}_{A}\cdot\hat{{\bf E}}^{\parallel}({\bf r}_{A})
  \simeq -[\hat{S}_{A23}\hat{{\bf E}}^{\parallel(+)}({\bf r}_{A})\cdot\hat{{\bf d}}_{A 23}+\mathrm{H.c.}]
\end{equation}
Combining (\ref{A4}), (\ref{A18}), and (\ref{A21}) gives
\begin{equation}
\label{A22}
\hat{H}_{AF}=-\hat{{\bf d}}_{A}\cdot\hat{{\bf E}}({\bf r}_{A})
  \simeq-[\hat{S}_{A23}\hat{{\bf E}}^{(+)}({\bf r}_{A})\cdot\hat{{\bf d}}_{A 23}+\mathrm{H.c.}]
\end{equation}
For three level atom, $k=1,2,3$, the atomic Hamiltonian (\ref{A10}) reads
\begin{equation}
\label{A23}
\hat{H}_{A}=\sum_{k} \hbar\omega_k \hat{S}_{Akk}=\hbar\omega_1 \hat{S}_{A11}+\hbar\omega_2 \hat{S}_{A22}+\hbar\omega_3 \hat{S}_{A33}
\end{equation}
Equations (\ref{A1}), (\ref{A2}), (\ref{A22}), and a subtraction of $\hbar(\omega_{2}-\omega_{0})\hat{I}$ from $\hat{H}_{A}$, Eq. (\ref{A23}) (taking into consideration that $\omega_{21}=\omega_{23}=\omega_{0}$), lead to the Hamiltonian 
\begin{equation}
\label{A24}
\hat{H}=\int d^{3}{\bf r}\int_{0}^{\infty} d\omega~\hbar\omega~ \hat{\bf f}^{\prime\dag}({\bf r},\omega)~\hat{\bf f}^{\prime}({\bf r},\omega)+\hbar\omega_{0}\hat{S}_{A22}-[\hat{S}_{A23}\hat{{\bf E}}^{(+)}({\bf r}_{A})\cdot\hat{{\bf d}}_{A 23}+\mathrm{H.c.}]
\end{equation}
\end{document}